\documentclass[12pt,preprint]{aastex}       

\newcommand{\Teff}{\mbox{$T_{\rm eff}$}}

\newcommand{\Lbol}{\mbox{$L_{\rm bol}$}}

\newcommand{\Lx}{{\mbox{$L_{\rm x}$}}/{\mbox{$L_{\rm bol}$}}~}

\newcommand{\lsun}{\mbox{$L_{\odot}$}}
\newcommand{\Mstar}{\mbox{$M_{\ast}$}}
\newcommand{\Rstar}{\mbox{$R_{\ast}$}}
\newcommand{\Msun}{\mbox{$M_{\odot}$}}
\newcommand{\Rsun}{\mbox{$R_{\odot}$}}

\newcommand{\Vinf}{\mbox{$v_{\infty}$}}
\newcommand{\vinf}{\mbox{$v_{\infty}$}}

\newcommand{\logg}{\mbox{$\log$~\textsl{g}}~}
\newcommand{\ha}{H$\alpha$}

\newcommand{\kms}{km s$^{-1}$}
\newcommand{\mdot}{$\dot M$}
\newcommand{\Mdot}{$\dot M$}
\newcommand{\myr}{M$_{\odot}$ yr$^{-1}$}
\newcommand{\Myr}{M$_{\odot}$ yr$^{-1}$}

\shorttitle{Effective Temperatures of Early-O Stars}
\shortauthors{Garcia \& Bianchi}

\begin{document}

\title{The Effective Temperatures of Hot Stars II.
The Early-O Types.
\footnote{
Based on observations  with the NASA-CNES-CSA
\it Far Ultraviolet Spectroscopic Explorer (\textit{FUSE}) \rm,
which is operated by The Johns Hopkins University under NASA contract NAS5-32985,
on \it International  Ultraviolet Explorer \rm
(\textit{IUE}) observations from the MAST and INES archives
and on MAST archival data from the \textit{Hubble Space Telescope} (\textit{HST})
and the \textit{Orbiting Retrievable Far and 
Extreme Ultraviolet Spectrometers} (\textit{ORFEUS}) mission.
}}
\author{ Miriam Garcia\altaffilmark{1} and Luciana Bianchi}
\affil{Center for Astrophysical Sciences \\
The Johns Hopkins University, Dept. of Physics and Astronomy \\
3400 N. Charles St., Baltimore,
MD 21218, USA.}
\email{garcia@pha.jhu.edu, bianchi@pha.jhu.edu}
\altaffiltext{1}{Departamento de Astrof\'{\i}sica, Universidad
de La Laguna, Avda. Astrof\'{\i}sico Francisco S\'anchez s/n,
38206 La Laguna (Tenerife), Spain. }

\begin{abstract}

We derived the stellar parameters of a sample
of Galactic early-O type stars by analysing their
UV and Far-UV spectra from
\textit{FUSE} (905-1187\AA),
\textit{IUE}, \textit{HST-STIS} and
\textit{ORFEUS} (1200-2000\AA).
The data have been modeled with spherical,
hydrodynamic, line-blanketed, non-LTE synthetic spectra 
computed with the \textit{WM-basic} code.
We obtain effective temperatures 
ranging from \Teff = 41,000~K to 39,000~K 
for the O3-O4 dwarf stars, and \Teff = 37,500~K
for the only supergiant of the sample (O4 If$^+$).
Our values are lower than those from previous empirical
calibrations for early-O types
by up to 20\%.
The derived luminosities of the dwarf stars are also
lower by 6 to 12\%;
however, the luminosity of the supergiant
is in agreement with previous calibrations
within the error bars.
Our results extend the trend found
for later-O types in a previous work by Bianchi \& Garcia. 

\end{abstract}


\section{INTRODUCTION}
\label{s_intro}

Hot massive stars have a great impact
on the surrounding interstellar medium (\textit{ISM}) and
play a crucial
role in the chemical evolution of galaxies. Their
strong ultraviolet radiation is responsible for the
ionization of nearby $HII$ regions where
their stellar winds blow vast bubbles. Their supersonic
wind outflows and the supernova explosion at the end of their evolution
transfer energy and momentum to
the \textit{ISM} and disperse the material processed in the
stellar interiors, thus setting the conditions for the
formation of subsequent generations of stars.

The determination of the physical parameters of massive 
stars is therefore of great interest, yet complicated. 
High resolution spectroscopy is needed.
Modeling the stellar atmosphere requires
to account  for the expanding wind, the
non local thermodynamic equilibrium (non-LTE) conditions
and the so called line-blanketing that modifies the
flux distribution, especially at short wavelengths.

Spectroscopy in the ultraviolet and far ultraviolet
ranges (hereafter UV and Far-UV) is a powerful
tool to study the winds of massive stars since
these spectral regions contain the resonance lines of
the most abundant ions in the wind. 
In this work we analyse spectra from the
\textit{Far Ultraviolet Spectroscopic Explorer}
(\textit{FUSE}) \citep[]{M00},
covering the 905-1187\AA\space region,
in conjunction with spectra at longer
wavelengths (1200-2000\AA) from
the \textit{International
Ultraviolet Explorer} (\textit{IUE}),
the \textit{Orbiting Retrievable Far and 
Extreme Ultraviolet Spectrometers} (\textit{ORFEUS})
and the 
\textit{Space Telescope Imaging Spectrograph} (\textit{STIS})
aboard the
\textit{Hubble Space Telescope} (\textit{HST}).
The \textit{FUSE} spectra allows us
to uniquely constrain the stellar parameters
by adding new ionization stages to those
accessible to \textit{IUE}, \textit{ORFEUS} and \textit{HST-STIS}
(e.g. \citet[]{LB00}, \citet[]{BG02}).

This is the second paper of a series devoted to
provide accurate and consistent determination  of the
stellar parameters of Galactic massive stars
with this method.
\citet[hereafter paper~I]{BG02} studied six mid-O type
stars and found effective temperatures
lower (by 15-20\%) than previously determined for the
sample stars or calibrated for their spectral types. In this work
we perform a similar analysis for early-O type stars.
The paper is organized as follows.
In Section \ref{s_data}
we provide details about the data and the reduction.
In Section \ref{s_stars} we summarize the relevant information
from the literature about the program stars. In Section \ref{s_morpho}
we compare the spectral line characteristics.
The stellar parameters are derived in Section \ref{s_models}
by modeling the spectra.
In Section
\ref{s_conclusion} the results are discussed.

\section{DATA AND REDUCTION}
\label{s_data}

For all the program stars we analysed
 \textit{FUSE} spectra (905-1187\AA)
and UV archival spectra ($\lambda>1200$\AA) from \textit{IUE}.
For a  few objects we also used \textit{ORFEUS} and
\textit{HST-STIS} archival data.
The datasets
used are listed in Table \ref{t_id}.

\begin{figure}
\epsscale{0.9}	
\plotone{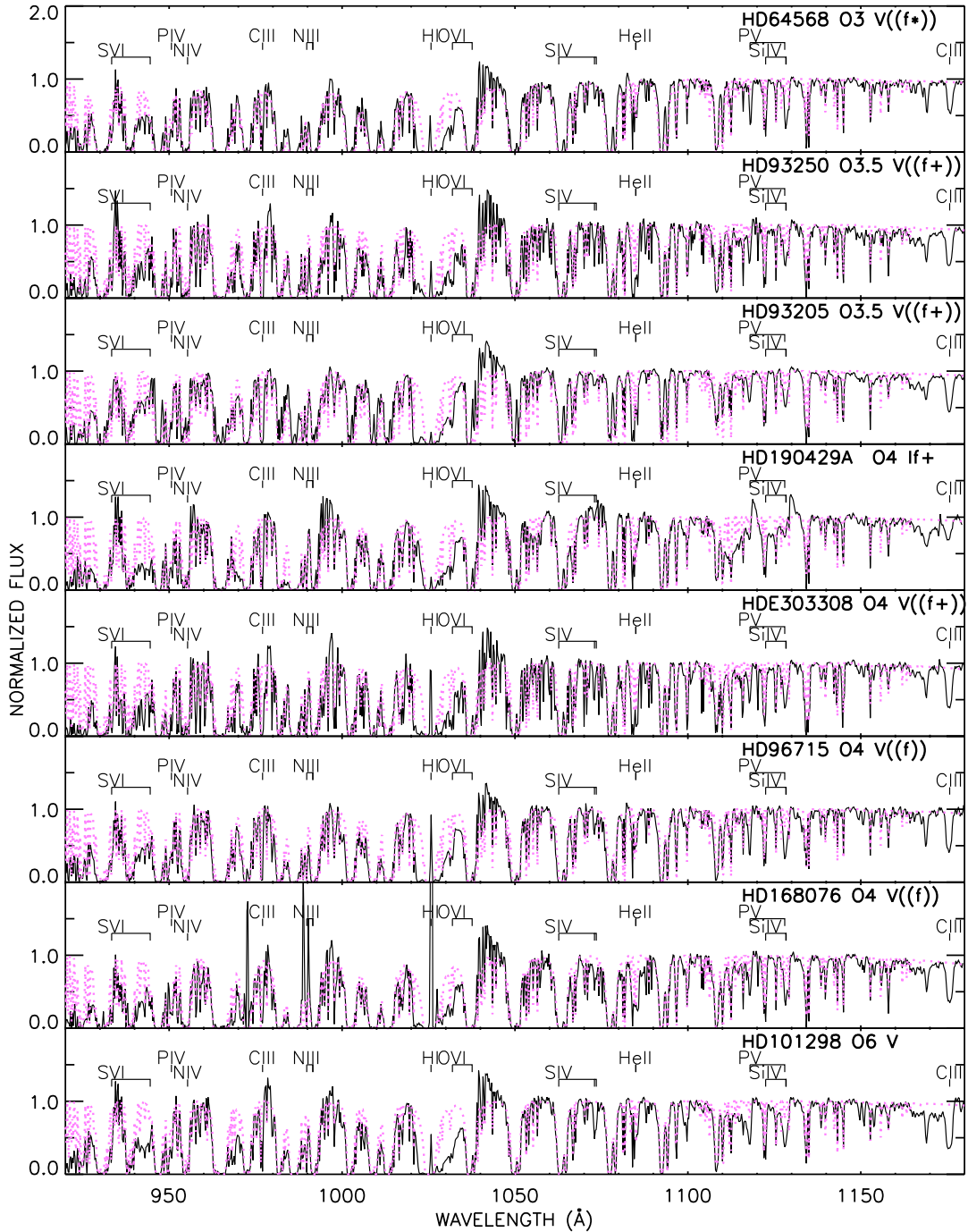}
 \caption[f1.eps]{
\scriptsize{
\textit{FUSE} spectra of the O3-O4 type stars analysed in this paper
and HD~101298 (O6~V, analysed in paper~I), arranged
by spectral type. The
spectra are rebinned to
steps of 0.25\AA\space for clarity and normalized
to the local continuum. The dotted (pink/light-grey) lines
are interstellar hydrogen ($H_{2}+HI$) absorption models,
calculated for the line of
sight of each star (see Section \ref{s_data}), that aid us
to identify the genuine
stellar features against the interstellar ones.
Note the stronger
profiles that 
\ion{P}{5}~$\lambda\lambda$1118.0,1128.0+\ion{Si}{4}~$\lambda\lambda$1122.5,1128.3,1128.4
and \ion{C}{4}~$\lambda$1169+\ion{C}{3}~$\lambda$1176 exhibit in the
spectrum of the supergiant star.
The dwarf stars have very similar spectral features,
but there is a smooth increase of
\ion{C}{3}~$\lambda$1176
and
\ion{P}{5}~$\lambda\lambda$1118.0,1128.0+\ion{Si}{4}~$\lambda\lambda$1122.5,1128.3,1128.4
from O3~V to O6~V.
}
 \label{f_all_fuse}}
\end{figure}

The \textit{FUSE} data, taken through
the \textit{LWRS} aperture (30\arcsec~ $\times$ 30\arcsec), 
have a resolution of $\lambda/\Delta\lambda \geq $ 20,000.
The data were processed with the pipeline (CalFUSE) version 2.0.5
\citep[]{pipeline}.
All the LiF and SiC segments were examined to assure optimal centering
of the spectra in the extraction window and to avoid
data defects such as event bursts or the ``worm'' 
\citep[]{Sahnow00,Sahnow02}.
The good portions from different channels were combined,
after the consistency of
wavelength scale and flux level was checked, to achieve the
maximum S/N. The wavelength ranges 905-930\AA\space and
1181-1187\AA, at the ends of the \textit{FUSE} range,
have poor spectral quality and were not used in this work.
The data in the 1082.5-1087\AA\space region come entirely
from the SiC channels, therefore
the \ion{He}{2}~$\lambda$1084.9 line 
was given a small weight in the analysis.
The resulting combined spectra,
normalized to the local continuum,
are shown in Figure \ref{f_all_fuse}.

\begin{figure}
\epsscale{1.15}
\plotone{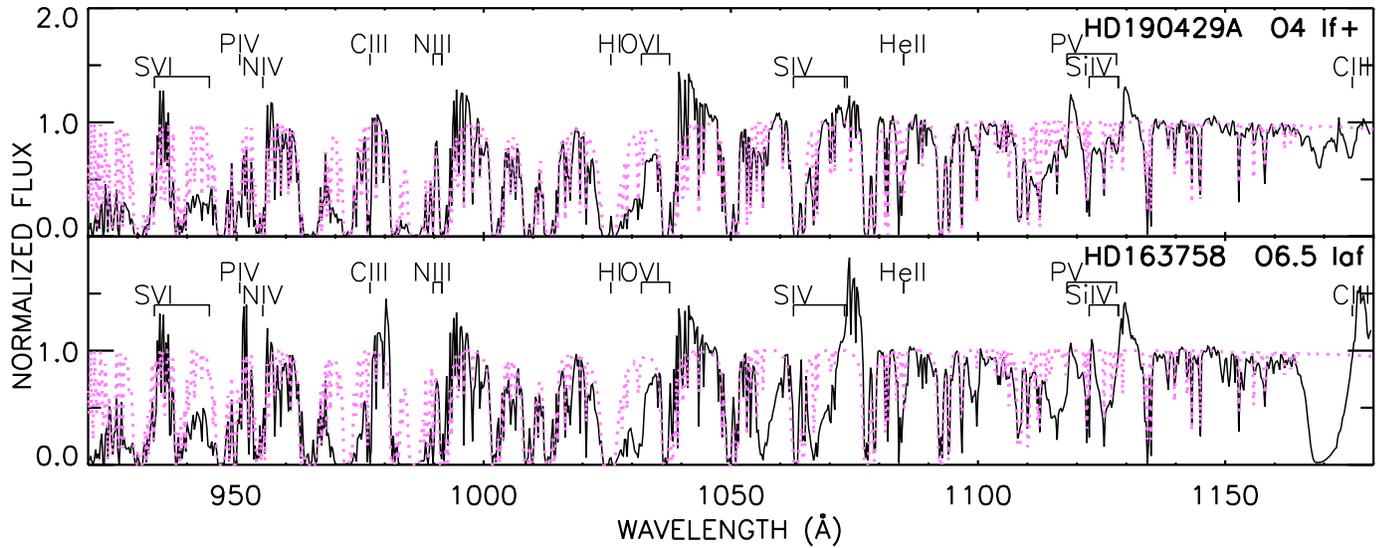}
 \caption[f3.eps]{
\footnotesize{
Temperature effects in luminosity class I:
\textit{FUSE} spectra of the supergiant stars
HD~190429A (O4~If$^+$)
and  HD~163758 (O6.5~Iaf, from paper~I)
(colors/lines as in Figure \ref{f_all_fuse}).
The strength of
\ion{S}{4}~$\lambda\lambda$1062.7,1073.0,1073.5
and
\ion{C}{4}~$\lambda$1169+\ion{C}{3}$\lambda$1176
severely decreases 
towards higher \Teff.
Note also that the wind profile of the
\ion{P}{5}~$\lambda\lambda$1118.0,1128.0+
\ion{Si}{4}~$\lambda\lambda$1122.5,1128.3,1128.4
blend changes in shape,
whereas the  \ion{O}{6}~$\lambda\lambda$1031.9,1037.6
P~Cygni profile is almost identical in both spectra.
}
 \label{f_2_superg_fuse}}
\end{figure}

\begin{figure}
\epsscale{0.8}
\rotatebox{180}{\plotone{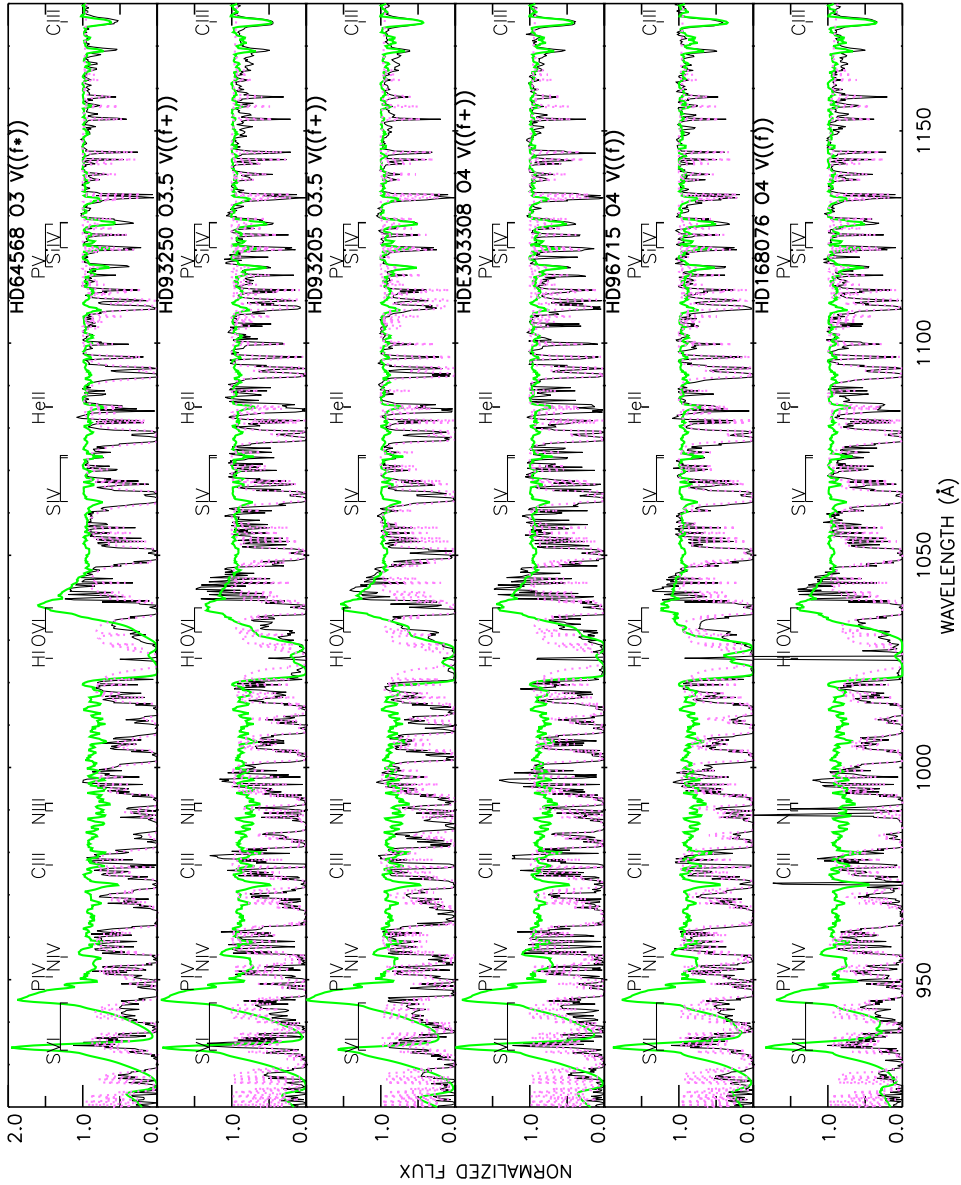}}
 \caption[f5.eps]{
{\footnotesize
Best fit \textit{WM-basic} models (green/grey)
and \textit{FUSE} spectra (black) of
the dwarf stars of the sample.
Models and spectra
have been rebinned to 0.25\AA~ steps.
The dotted (pink/light-grey) line is the interestellar
$H_{2}+HI$ absorption model in the line of sight of each star.
The \Teff~ of the models increases from
39,000~K (HD~160876, HD96715) to
40,000~K (HDE~303308, HD~93205 and HD~93250) and to
41,000~K (HD~64568). Note the good fit to
\ion{C}{3}~$\lambda$1176 and
\ion{P}{5}~$\lambda\lambda$1118.0,1128.0+
\ion{Si}{4}~$\lambda\lambda$1122.5,1128.3,1128.4,
the best indicators of temperature in this 
spectral range. 
}
\label{f_fit_all_fuse}
}
\end{figure}

The \textit{FUSE} wavelength range contains several absorption
lines and bands of interstellar atomic and molecular hydrogen.
In order to assess which features in
the spectrum are purely stellar and to determine reliable
continuum points for flux normalization, we calculated
the $H_{2}+HI$ absorption spectrum for the line of sight
of each star.
We used measurements of hydrogen column densities
when available in the literature or, otherwise, 
we estimated the column densitiy from color excesses
using the relations from  \citet[]{BO78}:
\textbf{ $N_H/E(\bv) = 5.8 \cdot 10^{21} \;
atoms \; cm^{-2} mag^{-1}$},
where $N_H=N_{HI}+2N_{H_2}$,
and 
\textbf{ $N_{HI}/E(\bv) = 4.8 \cdot 10^{21} \;
atoms \; cm^{-2} mag^{-1}$}.
The hydrogen ($HI+H_2$) absorption models are plotted
over the observed spectra in Figures \ref{f_all_fuse}, \ref{f_2_superg_fuse},
\ref{f_fit_all_fuse} and \ref{f_hd190429_col}.

\begin{figure}
\epsscale{0.9}  
\plotone{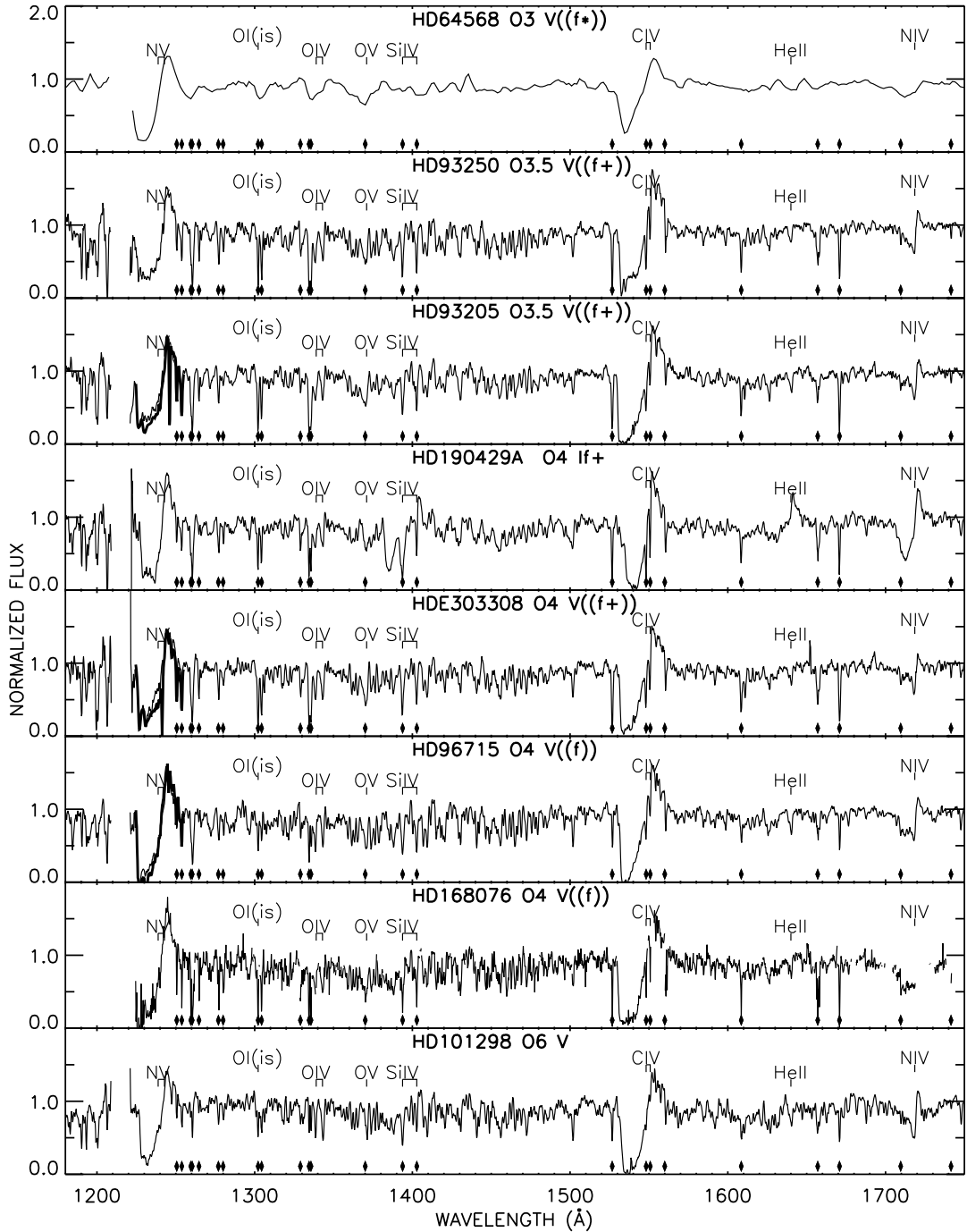}
 \caption[f2.eps]{
\scriptsize{Normalized
\textit{IUE} spectra of the stars included in Figure \ref{f_all_fuse}.
The spectra are rebinned
to  0.25\AA, 
except for the spectrum of HD~64568, which
has low resolution (6\AA).
The positions
of the most important
interstellar lines 
are marked with diamonds at the bottom
of the spectra. 
The \textit{IUE} data have very poor quality
in the  Ly$\alpha$ region, which has been removed
for clarity.
The saturated portions have been removed from the 
spectrum of HD~168076.
The thick lines plotted over the spectra
in the \ion{N}{5}~$\lambda\lambda$1238.8,1242.8 region
are \textit{ORFEUS-TUES} data for HD~96715
and \textit{HST-STIS} data for HDE~303308 and HD~93205.
Note that the \ion{Si}{4}~$\lambda\lambda$1393.8,1402.8
absorption weakens towards
earlier spectral types.
\ion{N}{4}~$\lambda\lambda$1718.0,1718.5 and
\ion{O}{5}~$\lambda$1371.0 and to a lesser extent,
\ion{C}{4}~$\lambda\lambda$1548.2,1550.8,
increase from the O6~V to the O3.5~V stars.
The most important luminosity effects are the stronger
\ion{He}{2}~$\lambda$1640.0,
\ion{Si}{4}~$\lambda\lambda$1393.8,1402.8 and
\ion{N}{4}~$\lambda\lambda$1718.0,1718.5
P~Cygni profiles of the supergiant spectra.
}
 \label{f_all_iue}}
\end{figure}

We examined all the existing  \textit{IUE} observations of the program objects
taken with the SWP camera (1150-1975\AA),
using the on-line tools of the MAST archive,
to check for variability and to exclude saturated portions, and
chose the data with the best S/N.
High dispersion \textit{IUE} spectra ($\Delta\lambda\approx$ 0.2\AA) are available
for all objects except for HD~64568, for which only a
low dispersion spectrum ($\Delta\lambda\approx$ 6\AA) exists.
We then downloaded the selected spectra
from the INES archive \citep[]{ines} because
the data typically have
a better background correction 
at wavelengths shorter than 1400\AA~
than the data in the MAST archive.
We additionally used
\textit{HST-STIS} spectra (0.2$\arcsec$x0.09$\arcsec$ aperture and E140H grating,
1150-1700\AA, $\lambda/\Delta\lambda$=114,000)
for two objects and
\textit{ORFEUS-TUES} spectra (900-1400\AA,
$\lambda/\Delta\lambda$=10,000)
for one program object. These data
were downloaded from the MAST archive.
The normalized
spectra in the UV range that contains the strongest
spectral lines
(1200-1750\AA) are shown in Figure \ref{f_all_iue}.

\section{THE PROGRAM STARS}
\label{s_stars}

We performed an exhaustive literature search 
about the program stars and collected all
data useful to this study.
In particular, we searched for reliable spectral classifications
and information about multiplicity and the environment of
the stars.
Table \ref{t_stars} compiles the spectral classification
and other relevant data, including the hydrogen column
densities in the line of sight of each object.
A discussion on the individual objects is given in the following
section.

\subsection{HD~190429A}
\label{s_hd190}
HD~190429 is a 
multiple system that belongs to the Cygnus OB3
association. HD~190429A has three
companions (B,C,D) located at 1.7$\arcsec$, 
42.5$\arcsec$ 
and 29.0$\arcsec$ 
respectively \citep[]{A86},
therefore only the B component is included 
in the \textit{FUSE} and \textit{IUE} large apertures.
The D component may be at the edge of the
\textit{FUSE} \textit{LWRS} slit, 
but it is considerably fainter than
HD~190429A ($\Delta$ V $\sim$ 4, \citet[]{A86}).
\citet[]{M98} reports a companion
of HD~190429A at a 0.09$\arcsec$ distance, but we found
no further information on this object.

For HD~190429A we adopt a spectral type of O4 If$^+$
\citep[]{W72,W73} and O9.5~II for the B component \citep[]{WH00}.
The luminosity class of the secondary
varies in the literature from Ibp \citep[]{morg55},
to III \citep[]{A86,CL74,guetter68}.
In Section \ref{s_190429} we estimate that its contribution 
to the total flux amounts to 15\% at most.
We examined the sixteen large aperture high resolution \textit{IUE} spectra
of HD~190429A
taken with the SWP camera.
Since we found no significant variation in
the flux levels and
the line profiles,
we include this object in our analysis.
The study of HD~190429A is of particular interest,
because it is the only supergiant star earlier than O5 
observed with \textit{FUSE} except for HD~93129A (O2If*), 
which also belongs to a binary system.

The \textit{IUE} spectrum of HD~190429A
shows characteristics intermediate between
O3~If$^*$ and O5~If$^+$ in
the UV morphological sequence
described by \citet[]{WN87}.
\citet[]{conti95} concur
that the UV spectral morphology of HD~190429A 
is typical of an O4-O5~If$^+$ star,
but found that its IR  spectrum (K-band) resembles 
that of a Wolf-Rayet,
indicating the presence of a strong wind.
\citet[]{morris96} obtained similar results.
\citet[]{WH00} found that the emission of \ha~ and
\ion{He}{2}$~\lambda$4686 in HD~190429A is stronger than
in other O3-O5~If$^+$ stars, again
suggesting that HD~190429A is evolving to  the WN stage.
However, the mass-loss rate we derive (Section \ref{s_models})
is consistent with that predicted by the
radiation pressure driven wind theory (Section \ref{s_conclusion}).

\subsection{HD~64568}

HD~64568 belongs to NGC~2467/Puppis~OB2 \citep[]{H72} and
is one of the ionizing stars of the irregular
$HII$ region Sh~2-311 \citep[]{SH2}. 
HD~64568 is a primary standard for the O3~V((f*)) type in the
recent revision of the spectral classification of early type objects
by \citet[]{W02a}, based on optical spectra.
We adopt this spectral classification,
although other works are discrepant:
\citet[]{mac76}, \citet[]{cram}, \citet[]{cruz74} and \citet[]{lo65}
classify HD~64568 as O5; \citet[]{mss} as O5/6 . None of these works provided
luminosity class. \citet[]{garr77} classified the star as O4~V ((f))
and \citet[]{P-J81} as O5~V.
The UV and Far-UV spectral morphology is consistent
with the O3~V((f*)) classification.

We did not find conclusive reports about binarity.
\citet[]{Sol86} analysed medium dispersion CTIO spectra
and found radial velocity variations, but could not determine
whether they are due to binarity
or to instabilities in
the atmosphere. \citet[]{cram72} obtained several measurements of the
radial velocity, and found a maximum variation of 10 \kms.

\subsection{HD~93250}

HD~93250 is located in the Carina Nebula
(NGC~3372), an $HII$ region
consisting of four lobes ionized by several stellar clusters,
including Trumpler~14
and Trumpler~16 (Tr~14 and Tr~16).
HD~93250 belongs to Tr~16.
We adopt the spectral type O3.5~V((f$^+$)) from
\citet[]{W02a}. 
Other authors agree that HD~93250 is a dwarf star
(O3~V((f)) \citep[]{W82,W71a}, O3:V((f)) \citep[]{LM82} and O3~V(f) \citep[]{TA74}).
However, HD~93250 is one of the brightest stars in
the Carina complex (V = 7.37, \citet[]{F82}).
\citet[]{C77} found evidence of a luminosity class higher than V
and classified the star as O3 (providing no luminosity class).
According to \citet[]{mathys88}, the star is a giant (O3~III(f))
and \citet[]{Petal96} classified it as a supergiant (O3~I)
based on its UV spectrum. We find that the FUSE spectrum
(see Figure \ref{f_all_fuse})
does not display the characteristic signatures of supergiants,
i.e. the \ion{S}{4}~$\lambda\lambda$1062.7,1073.0,1073.5
and  \ion{P}{5}~$\lambda\lambda$1118.0,1128.0
P~Cygni profiles
(see paper~I and Figures \ref{f_all_fuse}
and \ref{f_2_superg_fuse} of this paper),
supporting the luminosity class V.
On the other hand, the radius we derive
from line fitting (see Secton  \ref{s_93250})
is larger than for the other O3.5~V((f$^+$)) star of the sample,
consistently with the higher luminosity of HD~93250.

\citet[]{WP84} studied the UV-morphology of the early type dwarf
stars and found no anomaly in the \textit{IUE} spectrum of HD~93250,
except for the unsaturated profile of
\ion{C}{4}~$\lambda\lambda$1548.2,1550.8.
They suggested that an unresolved later-type companion could
be a possible explanation.
\citet[]{Lal91} measured a constant heliocentric radial velociy and
concluded that the star is not a binary; 
no companion was found in the speckle interferometric
survey of \citet[]{M98}.
We will further discuss this point in Section \ref{s_93250}.

\subsection{HD~93205}

HD~93205 is part of a multiple system, again located in the Tr~16 cluster.
HD~93205 is a spectroscopic binary 
with additionally a
visual companion (HD~93204, O5~V) 
at a 18.7\arcsec~ distance \citep[]{M98},
which may be included in the \textit{FUSE} \textit{LWRS} 
aperture, but not in the \textit{IUE} large aperture.
The primary component of the HD~93205
system is consistently classified
in the literature as O3~V (see Table \ref{t_stars})
but revised to O3.5~V((f$^+$)) by
\citet[]{W02a}, which we adopt.
\citet[]{CW76} and \citet[]{Mal01} provided spectral classification
for both components of the system: O3~V+O8~V.

The question whether HD~93205
is an eclipsing binary is not settled.
From the inclination angle (i $\simeq$ 45$\degr$-53$\degr$,
\citet[]{CW76}) no eclipses are expected.
\citet[]{Aal00} found photometric variability in the system
of a maximum of $\sim$ 0$^m$.02,
which they explained with a non uniform brightness distribution.
The latest photometric study \citep[]{vg03}
suggests instead that the light curve variations 
(amplitude 0$^m$.02)
are due to eclipses.
The only large aperture \textit{IUE} spectra, SWP14747 and SWP07959, were
taken at phases $\Phi$=0.78 and $\Phi$=0.28
(calculated using the orbital parameters 
from \citet[]{SL93});
the orbital positions of the secondary at these
phases are equivalent for the observer 
and, in fact, the spectrum does not vary.

The FUSE and IUE spectra of HD~93205
look very similar to those of HDE~303308
(which has a similar spectral type)
and the wind lines clearly originate from
the O3.5~V((f$^+$)) star, while the secondary
component (O8~V) may contribute to the continuum.
The spectral analysis
shows that this contribution is of the order
of $<$ 15\% (see Section \ref{s_dos}).

\subsection{HDE~303308}
HDE~303308, located 1$\arcmin$ North of $\eta$ Carinae,
also belongs to the Tr~16 cluster in the Carina Nebula.
We adopt the spectral classification of
\citet[]{W02a}, O4~V((f$^+$)), although the star 
has been previously classified as O3 dwarf
(see Table \ref{t_stars}).
Speckle interferometry
indicates that HDE~303308 is a single star \citep[]{M98}.
\citet[]{Lal91} report variability of the
radial velocity, but could not determine whether the star belongs
to a binary system.
\citet[]{SL01} measured the radial velocity
from three IUE spectra and found no significant changes.
\citet[]{vG89} did not find any  photometric variations.

\subsection{HD~96715}

HD~96715 belongs to the Carinae OB2 association,
rich in luminous massive stars. 
The spectral classifications throughout the
literature are in general agreement. We adopt 
O4~V((f)) \citep[]{W73,cruz74};
other similar classifications are listed 
in Table \ref{t_stars}.
HD~96715 is a blue straggler and displays an
unusually strong \ion{N}{3}~$\lambda$4514 line for 
its early type that may indicate nitrogen enrichment
\citep[]{SB86}.
There are no studies on binarity.

For this star we use a combination of two
large aperture \textit{IUE} spectra out of eight
available in the archive. We use SWP43980 for
$\lambda <$ 1700\AA, which has the best S/N ratio
in this spectral region, but it is saturated
for $\lambda >$ 1700\AA\space where we use SWP21999.

\subsection{HD~168076}

HD~168076 is a member of the young open cluster NGC~6611 \citep[]{W61},
located near the outermost part of the Sagittarius-Carina spiral arm.
The spectral classification we adopt is O4~V((f))
\citep[]{W73}. Other authors provide similar
classifications (see Table \ref{t_stars}) except for
\citet[]{mathys88}: O4~III(f) and
\citet[]{ngc6611}: O5~V((f$^{*}$)).
\citet[]{rachford} studied the line of sight of HD~168076
from its \textit{FUSE} spectra
and calculated the column densities of $HI$ and $H_2$
(listed in Table \ref{t_stars}) which we use in this analysis.

HD~168076 is a visual binary. \citet[]{Dal01}
resolved the components with a high angular resolution
(0.035$\arcsec$) adaptive optics system and estimated a
separation of $\sim$ 0.15\arcsec~ and a  difference
in magnitude of $\Delta$K=1.57.
However, the line spectrum of HD~168076, very similar to the 
spectra of the other O4 dwarf stars in the sample,
is apparently dominated in the Far-UV and UV ranges by the 
lines from the hot component.

There is only one high dispersion \textit{IUE} spectrum of this object. It has some
saturated portions, removed from
the plots in Figures \ref{f_all_iue} and \ref{f_fit_all_iue}
and not considered in the analysis.

\section{ANALYSIS OF THE SPECTRAL MORPHOLOGY}
\label{s_morpho}
Examination of the UV and Far-UV spectra
of the sample stars
reveals the correspondence
of the line morphology in this spectral region with
the spectral classifications derived
from the optical range.
The general behaviour of the spectral lines as a function of
spectral type and luminosity class 
has been described in a number of atlases, both
in the \textit{IUE} range \citep[]{IUEatlas} 
and in the \textit{FUSE} range (\citet[]{MWatlas} for Galactic stars,
\citet[]{Wal02b} for Magellanic Cloud stars).
In this section we provide a more detailed discussion of the spectral
morphology within the spectral types covered in this
work and paper~I. 
In the quantitative spectral modeling
(Section \ref{s_models})
the observed line variations will be
 explained in terms of physical parameters.

\subsection{Luminosity Effects at O4}
\label{s_O4}

The \textit{FUSE} and \textit{IUE} spectra of the 
O4 type stars are shown in Figures \ref{f_all_fuse} and
\ref{f_all_iue}.
The sample consists of one supergiant star
(HD~190429A) and three dwarf stars (HDE~303308,
HD~96715 and HD~168076). The \textit{FUSE} satellite has
not observed any giant star of type O4 (or earlier) yet.

The most prominent variations among luminosity classes are the
different strength of \ion{P}{5}~$\lambda\lambda$1118.0,1128.0
in the \textit{FUSE} range and \ion{Si}{4}~$\lambda\lambda$1393.8,1402.8,
\ion{He}{2} $\lambda$1640 and
\ion{N}{4}~$\lambda\lambda$1718.0,1718.5
in the \textit{IUE} range. 
\ion{P}{5}~$\lambda\lambda$1118.0,1128.0,
\ion{Si}{4}~$\lambda\lambda$1393.8,1402.8 and
\ion{He}{2} $\lambda$1640 display
wind profiles in the spectrum of the supergiant star but are photospheric
absorptions in the dwarf stars.
\ion{S}{4}~$\lambda\lambda$1062.7,1073.0,1073.5,
which shows a P~Cygni profile in HD~190429A
(though severely masked by interstellar hydrogen absorption),
and \ion{C}{4}~$\lambda$1169+\ion{C}{3}$\lambda$1176
change similarly. 
The behaviour of these lines
is similar to what we found for mid-O types (paper~I)
and it is caused by the 
mass-loss rate variation with luminosity.
In the present sample, the
\ion{S}{4}~$\lambda\lambda$1062.7,1073.0,1073.5 and
\ion{C}{4}~$\lambda$1169+\ion{C}{3}$\lambda$1176
profile changes are not as remarkable as in later types
because these lines are much less conspicuous
due to the higher \Teff.

Note that \ion{O}{6}~$\lambda\lambda$1031.9,1037.6
displays a well developed P~Cygni profile of similar
strength at all luminosity classes as
already seen in mid-O type stars (paper~I).
Similarly, the lines of other high ionization species,
\ion{N}{5}~$\lambda\lambda$1238.8,1242.8 and
\ion{C}{4}~$\lambda\lambda$1548.2,1550.8, do not vary
appreciably because they are saturated.

\subsection{Temperature Effects}
\label{teff_effects}

We find  systematic line variations 
when we compare
the spectra of the O3-O4 type stars
analysed in this paper to the later
type stars from paper~I.
Despite their many similarities, the 
spectra of O3-O6 dwarf stars in Figures  \ref{f_all_fuse}
and \ref{f_all_iue} display  clear  \Teff~ effects.
Notably,
\ion{C}{4}~$\lambda$1169+\ion{C}{3}$\lambda$1176
and
\ion{P}{5}~$\lambda\lambda$1118.0,1128.0+\ion{Si}{4}~$\lambda\lambda$1122.5,1128.3,1128.4
decrease towards earlier types.
In the \textit{IUE} range,
the weaker \ion{Si}{4}~$\lambda\lambda$1393.8,1402.8 and
stronger \ion{N}{4}~$\lambda\lambda$1718.0,1718.5,
\ion{O}{5}~$\lambda$1371.0 and
\ion{C}{4}~$\lambda\lambda$1548.2,1550.8
also indicate higher \Teff.

The line variations with spectral type are
more remarkable in the supergiants, as wind
features are more developed in their spectra.
When we compare the \textit{FUSE} spectra of
O4 and O6.5 supergiants
(see Figure \ref{f_2_superg_fuse})
we find the latter to have stronger
\ion{S}{4}~$\lambda\lambda$1062.7,1073.0,1073.5
and 
\ion{C}{4}~$\lambda$1169+\ion{C}{3}$\lambda$1176,
whereas
\ion{O}{6}~$\lambda\lambda$1031.9,1037.6
does not vary.
\ion{P}{5}~$\lambda\lambda$1118.0,1128.0
has a P~Cygni
profile in both stars, however with different shapes,
indicating 
a different ionization structure in the wind.

\begin{figure}
\epsscale{1.15}
\plotone{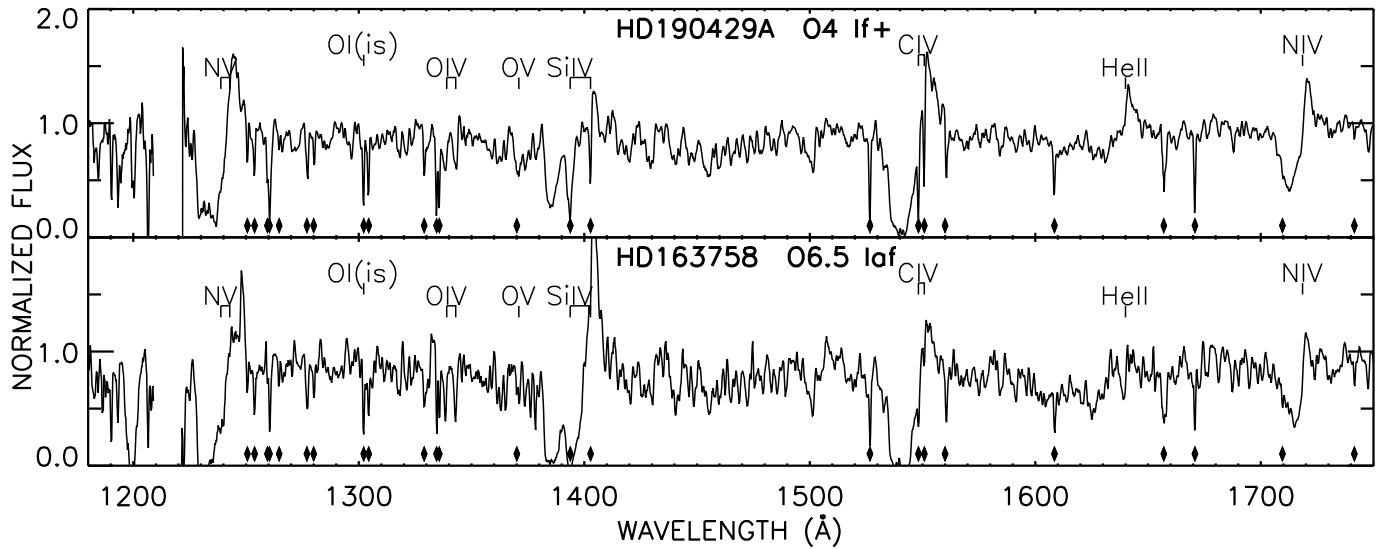}
 \caption[f4.eps]{
\footnotesize{
\textit{IUE} spectra of HD~190429A and HD~163758
(lines/symbols as in Figure \ref{f_all_iue}).
The most outstanding differences are seen in the
\ion{Si}{4}~$\lambda\lambda$1393.8,1402.8 doublet
and
the \ion{He}{2}~$\lambda$1640.0 line.
\ion{N}{4}~$\lambda\lambda$1718.0,1718.5 and
the emission of
\ion{C}{4}~$\lambda\lambda$1548.2,1550.8
are stronger in the earlier type star.
}
 \label{f_2_superg_iue}}
\end{figure}

In the \textit{IUE} range (see Figure \ref{f_2_superg_iue}),
the lines showing the largest variation
are \ion{Si}{4}~$\lambda\lambda$1393.8,1402.8
and \ion{He}{2}~$\lambda$1640. The
strength of
\ion{Si}{4}~$\lambda\lambda$1393.8,1402.8
drastically decreases from
HD~163758,
where it is a fully developed P~Cygni
profile, to HD~190429A.
The spectrum of HD~190429A displays a
strong emission
of \ion{He}{2}~$\lambda$1640.0, characteristic
of O3-O5~If stars
\citep[]{IUEatlas,WN87}.
\ion{N}{4}~$\lambda\lambda$1718.0,1718.5
and the emission of
\ion{C}{4}$\lambda\lambda$1548.2,1550.8
increase from type O6.5 to O4.
The strength of the feature at 1500\AA,
which we believe  to be a \ion{S}{5} line
possibly contaminated with \ion{Si}{3} lines
(see paper~I), is approximately the same in both spectra.
The absorption line
of \ion{O}{5}~$\lambda$1371.0 is present
only in the spectra of the hotter star, HD~190429A.

\section{QUANTITATIVE SPECTRAL MODELING}
\label{s_models}

In this section
we determine the photospheric and wind parameters
of the sample stars by fitting their
Far-UV and UV spectra 
with spherical, hydrodynamic, line-blanketed,
non-LTE synthetic models. The models
were calculated with the version 2.11 of the \textit{WM-basic} code
\citep[]{WMBAS},
but have the 'solar' abundances of the previous 
version (\textit{WM-basic} 1.22)
for consistency with the grid of models constructed for paper~I.

Our study takes into account the effects of shocks in the wind.
Shocks produce soft-X
rays in the expanding atmosphere which affect the ionization
of several species, most remarkably \ion{O}{6} and \ion{N}{5}. By
modeling the lines of highly ionized atoms in
the \textit{FUSE} and \textit{IUE} ranges
consistently with the other
spectral features, we can determine the value of the parameter
\Lx,
and thus provide a unique solution for the stellar
parameters (\Teff, \logg, \Rstar, \vinf~ and \Mdot).

We have built a vast grid of \textit{WM-basic} models
which -by comparison to the observed spectra- provides upper and
lower limits of the stellar parameters. The
final values 
and their
uncertainties are obtained by computing models that
refine the grid within the range of interest
for each observed spectrum.
In the following section
we provide further details of the fitting process.
The best fit models
are shown in Figures \ref{f_fit_all_fuse}-\ref{f_hd190429_col}. The
derived stellar parameters are compiled in Table \ref{t_models}.

\subsection{The Dwarf Stars}
\label{s_fit_v}

The spectra of the dwarf stars are mostly similar,
except for subtle line variations
(see Section \ref{teff_effects}). 
Therefore we 
follow a similar procedure to fit their spectra,
which we explain in this section.
The analysis of HD~93250 and HD~64568 is slightly different and
the details are given in Sections \ref{s_93250}
and \ref{s_64568}.

We learned from our grid of models that
\ion{P}{5}~$\lambda\lambda$1118.0,1128.0,
\ion{C}{4}~$\lambda$1169+\ion{C}{3}~$\lambda$1176,
\ion{O}{5}~$\lambda$1371.0 and
\ion{N}{4}~$\lambda\lambda$1718.0,1718.5 are
good temperature indicators in the interval of \Teff~ of
38,000~K to 42,000~K (enclosing the temperatures 
of our sample as we will see) and mass-loss rates of
$10^{-7}$ to $10^{-6}$ \myr;
we initially assumed
\logg= 4 and \Rstar= 9\Rsun, adequate
for luminosity class V.
In this range of parameters,
\ion{P}{5}~$\lambda\lambda$1118.0,1128.0 and
\ion{C}{4}~$\lambda$1169+\ion{C}{3}~$\lambda$1176
decrease with temperature; both lines are
photospheric and therefore
insensitive to changes of \Mdot~ and \Lx.
The strength of \ion{N}{4}~$\lambda\lambda$1718.0,1718.5
increases with \Teff~ and \Mdot~ and marginally decreases
with higher \Lx.
\ion{O}{5}~$\lambda$1371.0 has a hard threshold,
and forms only in models with \Teff$\ge$ 40,000~K
(independently of shocks and mass-loss rate)
thus providing an upper limit to the temperature
of stars not displaying this line.
When  \ion{O}{5}~$\lambda$1371.0 is present 
in the stellar spectrum,
its strength varies similarly to 
\ion{N}{4}~$\lambda\lambda$1718.0,1718.5
and helps us constrain \Lx and \Mdot.

We can set solid limits to the interval of possible
temperatures of the dwarf stars, based on the 
strength of these lines in their spectra.
HD~64568, HD~93250, HD~93205 and HDE~303308
must have a temperature higher or equal than 
40,000~K since their spectra display the
\ion{O}{5}~$\lambda$1371.0 line.
For these stars, the upper limit to \Teff~ is
42,000~K, because at higher temperatures
(in the range of \Mdot~ set above),
\ion{O}{5}~$\lambda$1371.0 becomes a fully developed
P~Cygni profile in the models, not seen in 
any stellar  spectra,
and \ion{N}{4}~$\lambda\lambda$1718.0,1718.5
is not as strong as observed.
\ion{O}{5}~$\lambda$1371.0 is absent in
the spectra of HD~96715 and HD~168076, thus their
temperature must be below 40,000~K.
The lower limit to the temperature of these stars
is established with \ion{P}{5}~$\lambda\lambda$1118.0,1128.0
and secondarily \ion{C}{4}~$\lambda\lambda$1548.2,1550.8.
\ion{P}{5}~$\lambda\lambda$1118.0,1128.0
displays a wind profile (discrepant with the observed
photospheric profile)
in the models with \Teff$\leq$37,000~K.
In the 38,000~K to 42,000~K range all our models have
an excess emission of
\ion{C}{4}~$\lambda\lambda$1548.2,1550.8
(with respect to the observations)
but successfully fit
the observed absorptions;
however, at \Teff$\leq$37,000~K the profile of the 
absorption changes (for any
mass-loss rate) and does not fit the spectrum.

The \Teff~ of each object is determined from
the best fit to the
\ion{C}{4}~$\lambda$1169+\ion{C}{3}~$\lambda$1176
and \ion{P}{5}~$\lambda\lambda$1118.0,1128.0
lines. We find \Teff=41,000~K for HD~64568, \Teff=40,000~K
for HD~93250, HD~93205 and HDE~303308
and \Teff=39,000~K for HD~96715 and HD~168076.
The error bars given
in Table \ref{t_models} ($\pm$ 2,000~K) are larger than
the range of acceptable temperatures of each object indicated
by the spectra, because they also account for systematic
errors, such as the uncertainty introduced by the flux
normalization and the possible contribution by
companions ($<$ 15\%, see later) in some cases.

Once we delimit the effective temperature,
we proceed to derive \Mdot.
Most of the lines useful for this purpose
in the analysis of mid-O type stars
(\ion{P}{5}~$\lambda\lambda$1118.0,1128.0,
\ion{C}{4}~$\lambda$1169+\ion{C}{3}~$\lambda$1176,
\ion{N}{5}~$\lambda\lambda$1238.8,1242.8 and
\ion{C}{4}~$\lambda\lambda$1548.2,1550.8) are insensitive to
the variation of \Mdot~ at the  high temperatures
of the early-O stars. The mass
loss rate is then determined
with the aid of the less prominent
\ion{O}{5}~$\lambda$1371.0,
\ion{N}{4}~$\lambda\lambda$1718.0,1718.5,
and \ion{Si}{4}~$\lambda\lambda$1393.8,1402.8
(increasing with higher \Mdot),
and secondarily with
\ion{O}{6}~$\lambda\lambda$1031.9,1037.6 (decreasing with higher \Mdot).
For each object, the lower/upper limits to mass-loss rate
are set to the values that produce
\ion{O}{5}~$\lambda$1371.0 and 
\ion{N}{4}~$\lambda\lambda$1718.0,1718.5
in defect/excess.

The most sensitive lines to \Lx are
\ion{O}{6}~$\lambda\lambda$1031.9,1037.6 and
\ion{N}{5}~$\lambda\lambda$1238.8,1242.8,
as already noted in paper~I.
\ion{S}{6}~$\lambda\lambda$933.4,944.52 is also
sensitive, but it is given little weight in the fitting
process as this region is heavily affected by the interstellar
hydrogen absorption. 
\ion{N}{5}~$\lambda\lambda$1238.8,1242.8
presents a problem related to its unsaturated absorption
(which we discuss in detail below)
and cannot be used to constrain the shocks.
\Lx is then derived from the model that provides the
best fit to \ion{O}{6}~$\lambda\lambda$1031.9,1037.6,
which increases with increasing shocks.
In the range of temperatures and mass-loss rates
we found for the dwarf stars,
\Lx must be comprised between -7.5
(minimum to reproduce the observed
\ion{O}{6}
profile) and -6.5 (when this line
is in excess in the models, independently
of the other stellar parameters),
but can be further constrained
for each object once \Teff~ and \Mdot~
are known (as reflected in Table \ref{t_models}).

We then calculated models with different radii 
(in the range 6-12 \Rsun)
for \logg=4.0.
We find that if we increase
the radius (i.e. we increase  mass and luminosity), 
\ion{O}{6}~$\lambda\lambda$1031.9,1037.6 increases while
\ion{P}{5}~$\lambda\lambda$1118.0,1128.0,
\ion{C}{3}~$\lambda$1176 and
\ion{N}{4}~$\lambda\lambda$1718.0,1718.5 decrease. At
\Rstar=6\Rsun, the models do not reproduce the P~Cygni profile of
\ion{O}{6}. A model with a radius of \Rstar=12\Rsun~ provides an
overall worse fit to the spectra of HD~96715, HD~168076,
HDE~303308 and HD~93205 than \Rstar=9\Rsun. 
For these stars we adopt a radius of 
\Rstar=9\Rsun~ $\pm$ 3 \Rsun~ (HD~64568
and HD~93250 are discussed in Sections \ref{s_93250}
and \ref{s_64568}).

There is a limited number of photospheric
lines sensitive to gravity
in the FUSE and IUE ranges and, by consequence,
the derived value of \logg is less well
constrained than the other stellar parameters.
\ion{P}{5}~$\lambda\lambda$1118.0,1128.0 and
\ion{N}{4}~$\lambda\lambda$1718.0,1718.5 increase with
decreasing gravity, and \ion{C}{3}~$\lambda$1176 has the
opposite behavior. 
However, the strength of these lines
depends on \logg more weakly than on \Teff~ and
\Mdot. We find
that models with \logg=3.8 or 4.2 give poorer fits
than \logg=4.0 and we adopt these values as the error margin.
Our findings agree with the results from works
based on analyses of optical spectra
that derive \logg for HD~93205 and HDE~303308
between 3.9 and 4.1 (see Table \ref{t_prev})
and with the calibration of \logg with spectral
type of \citet[]{MPRM03} and \citet[]{vacca} that assign
\logg$\simeq$3.9 to O3~V and O4~V stars.
For HD~168076, \citet[]{M75} derived \logg=4.3
from photometric data.

\begin{figure}
\epsscale{0.8}
\rotatebox{180}{\plotone{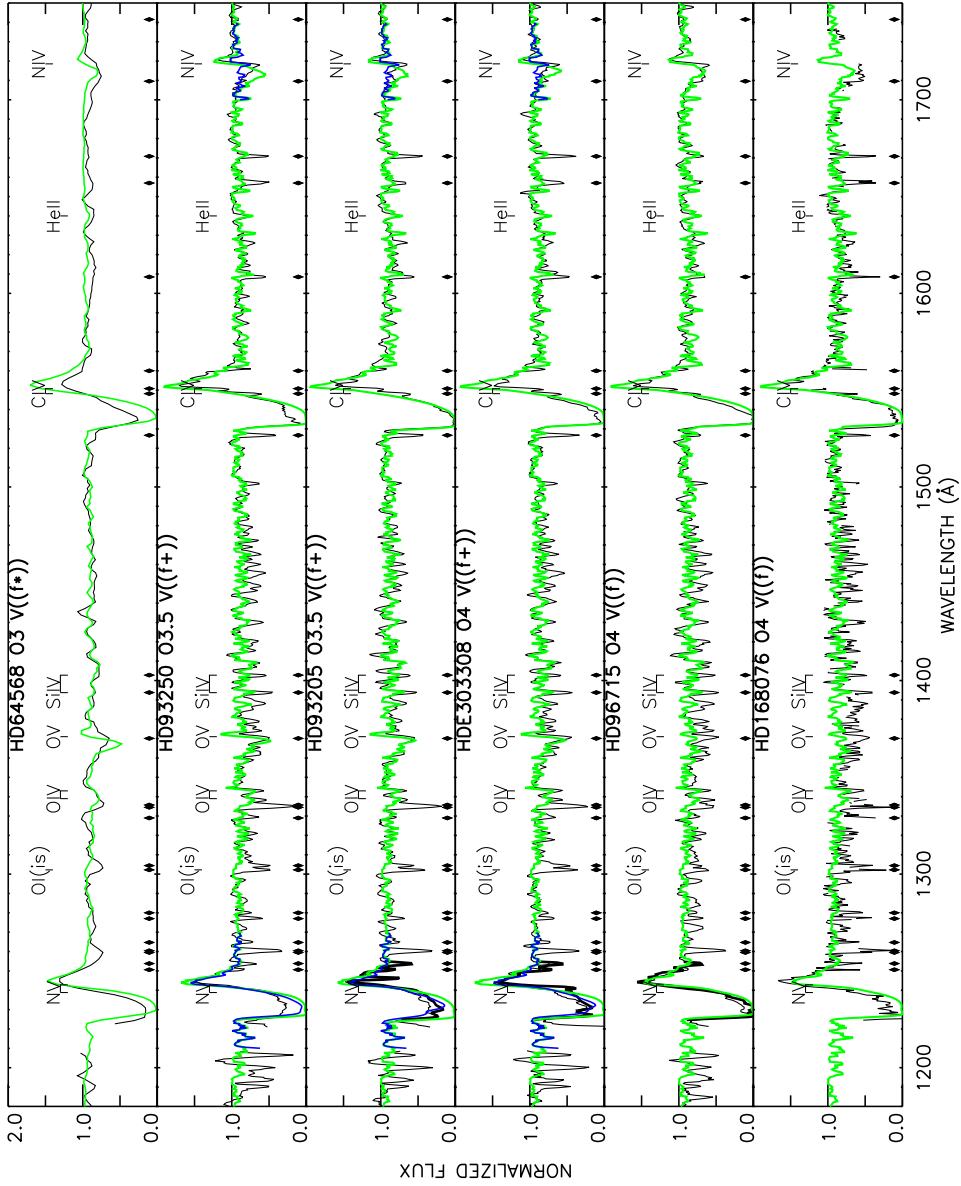}}
 \caption[f6.eps]{
{\footnotesize
Same as Figure \ref{f_fit_all_fuse}, IUE range.
For HD~93250, HD~93205 and HDE~303308 we also include in
blue/dark-grey
the best fit model with an underabundance
of nitrogen ($\epsilon _N = 0.1 \; \epsilon _{N,\odot}$)
in the spectral regions around
\ion{N}{5}~$\lambda\lambda$1238.8,1242.8
and
\ion{N}{4}~$\lambda\lambda$1718.0,1718.5
(the rest of the model is identical
to that with solar abundances).
The observed and synthetic
spectra have been rebinned to 0.5\AA~ steps, except
for HD~64568 
(model rebinned to 6\AA).
Other lines and symbols have the same meaning as in
Figure \ref{f_all_iue}. 
The best \Teff~ indicators in this range are
\ion{N}{4}~$\lambda\lambda$1718.0,1718.5
and \ion{O}{5}~$\lambda$1371.0,
which we successfully fit for all stars but
HD~168076.
The unsaturated 
\ion{N}{5}~$\lambda\lambda$1238.8,1242.8
profiles are discussed in the text.
The model fits well the
\textit{ORFEUS} spectrum of HD~96715.
}
\label{f_fit_all_iue}
}
\end{figure}

The best fit models 
are shown in Figures \ref{f_fit_all_fuse} 
and \ref{f_fit_all_iue}.
The synthetic spectra reproduce with great accuracy most of the
UV and Far-UV lines. Yet, they fail to reproduce the
\ion{N}{5}~$\lambda\lambda$1238.8,1242.8 
unsaturated profile in the \textit{IUE}
spectra of all stars.
The \textit{IUE} data may suffer from
inaccurate background subtraction
in this wavelength range due to the
echelle order overlap \citep[]{bibo84}, so
we searched for other UV spectra
in the MAST archive.
In Figures \ref{f_all_iue} and \ref{f_fit_all_iue} 
we have included data in the
\ion{N}{5}~$\lambda\lambda$1238.8,1242.8
region from \textit{ORFEUS-TUES} for HD~96715
and from \textit{HST-STIS} for HDE~303308 and HD~93205.
The \textit{ORFEUS-TUES} spectrum of HD~96715
displays a saturated \ion{N}{5}~$\lambda\lambda$1238.8,1242.8
P~Cygni profile
that the \textit{WM-basic} model fits well,
therefore at least in this case the mismatch 
is indeed likely to originate from the IUE data reduction.
However, \ion{N}{5}~$\lambda\lambda$1238.8,1242.8 
is unsaturated in the \textit{STIS} spectra of
both HDE~303308 and HD~93205.
While the latter is a binary, thus the residual flux
may originate from the cooler companion, the profile
is hard to explain in HDE~303308.
This prompted us to explore 
the grid of \textit{WM-basic} models,
to check if the combination of parameters
that produces unsaturated
\ion{N}{5}~$\lambda\lambda$1238.8,1242.8
overlaps with the range of parameters
we found adequate for early type dwarf stars.

The behaviour of the \ion{N}{5}~$\lambda\lambda$1238.8,1242.8 doublet 
in models with \logg~= 4 and \Rstar~= 9\Rsun~
can be summarized as follows.
It becomes stronger with increasing
temperature up to $\sim$ 44,000~K
(the exact value depends on \Mdot~ and \Lx)
and then it decreases for \Teff$\gtrsim$ 44,000~K.
The doublet
grows with increasing shocks up to $\log$\Lx$\leq$-6.75,
then it decreases for $\log$\Lx$\geq$-6.75.
In the range of temperatures we have determined for
O3-4~V stars (38,000-42,000~K), 
higher \Mdot~ produces a stronger
\ion{N}{5}~$\lambda\lambda$1238.8,1242.8 profile.

If we decrease \Teff~ so that the  models 
produce unsaturated \ion{N}{5}~$\lambda\lambda$1238.8,1242.8,
we canot reproduce
the observed \ion{O}{6}~$\lambda\lambda$1031.9,1037.6.
If instead we increase \Teff, we achieve 
unsaturated 
\ion{C}{4}~$\lambda\lambda$1548.2,1550.8
at lower \Teff~ (by  2,000~K) than needed to produce
unsaturated
\ion{N}{5}~$\lambda\lambda$1238.8,1242.8
(unsaturated \ion{C}{4}~$\lambda\lambda$1548.2,1550.8
is in disagreement with the observed
spectra of all dwarf stars except for HD~93250 and HD~64568).
Reduced shocks underestimate
\ion{O}{6}~$\lambda\lambda$1031.9,1037.6.
Model spectra calculated with lower mass-loss rate do not
have \ion{N}{4}~$\lambda\lambda$1718.0,1718.5.
We can fit \ion{N}{5}~$\lambda\lambda$1238.8,1242.8
in the spectra of HD~93205 and HDE~303308
(but not HD~93250) with a
nitrogen abundance of
$\epsilon _N = 0.1 \; \epsilon _{N,\odot}$
but the models lack
\ion{N}{4}~$\lambda\lambda$1718.0,1718.5
(see Figure \ref{f_fit_all_iue}).
Note that none of the stars
analysed in this work is in the OBC/OBN compilation
of \citet[]{W76},
nor we found any publication that measured
non-solar compositions for them.

In conclusion, we cannot compute a \textit{WM-basic}
model that displays unsaturated\\
\ion{N}{5}~$\lambda\lambda$1238.8,1242.8
and is in agreement
with the rest of the observed spectrum.
We suspect that a shock (and X-ray ionization)
treatment different from what used by WM-basic
may reproduce the
\ion{N}{5}~$\lambda\lambda$1238.8,1242.8 P~Cygni
profile.
We will investigate this possibility by
comparing models computed with WM-basic
and CMFGEN (\citet[]{BGH03}, Bianchi, Garcia, \& Herald, in preparation).

\subsubsection{HD~93205 and HD~168076}
\label{s_dos}
There is an additional source of uncertainty in 
the analysis of HD~93205 and HD~168076.
Both of them are binaries whose secondary component
is most likely included in the FUSE and IUE apertures,
its contribution to the
observed total flux being however hard to quantify.

The secondary component of HD~93205 is
an O8~V star. 
From our grid of synthetic spectra
we know that the O8~V
continuum flux (\Teff=30,000~K, from Figure \ref{tefflum}) can
be up to 50\% the flux of an O3.5~V 
(\Teff=40,000~K, this work) in the
900-1800\AA~ range if the stars had the same radius.
Yet, we can narrow down this upper limit.
\ion{N}{5}~$\lambda\lambda$1238.8,1242.8 is almost
absent in the UV spectrum of O8~V stars
\citep[]{IUEatlas}.
If we assume that this line is saturated in O3.5~V stars
(see, for example, the \textit{ORFEUS} spectrum of HD~96715
in Figure \ref{f_all_iue})
the residual flux at its core
in the IUE spectrum, of $\sim$ 15\%,
may originate from the secondary and/or
inaccurate background subtraction, as explained previously.
As for other lines, the spectrum of an O8~V star
displays photospheric profiles of
\ion{Si}{4}~$\lambda\lambda$1393.8,1402.8
and \ion{N}{4}~$\lambda\lambda$1718.0,1718.5,
and unsaturated
\ion{C}{4}~$\lambda\lambda$1548.2,1550.8
\citep[]{IUEatlas}.
This could explain the subtle mismatch of the best fit model
to the observed emission of
\ion{C}{4}~$\lambda\lambda$1548.2,1550.8
and the absorptions of 
\ion{Si}{4}~$\lambda\lambda$1393.8,1402.8
and \ion{N}{4}~$\lambda\lambda$1718.0,1718.5.

The HD~168076 system is much less known.
Since we do not know the spectral type of the secondary
component, we cannot speculate the contamination to the
O4 V((f)) star UV and Far-UV spectra. 
These, however, are very similar to the spectra of the other
dwarves in the sample, therefore the contribution from the 
secondary cannot be conspicuous.
A possible effect 
from the secondary is that the
\ion{N}{5}~$\lambda\lambda$1238.8,1242.8 and
\ion{C}{4}~$\lambda\lambda$1548.2,1550.8 
features are not saturated.
Again, if we suppose that these lines are saturated in the
spectrum of the main component,
we can estimate the contribution of the secondary to be
$\lesssim$ 10\% to the total flux in the \textit{IUE} range.

\subsubsection{\it HD~93250} 
\label{s_93250}
The \textit{IUE} spectrum of HD~93250 displays an
unsaturated \ion{C}{4}~$\lambda\lambda$1548.2,1550.8
P~Cygni profile.
This feature was noted by \citet[]{IUEatlas} 
who suggested that it may
indicate a hidden companion. 
Yet, the star has not been found to be binary
(see Section \ref{s_stars}).
A poor background subtraction in the IUE spectrum may make
\ion{N}{5}~$\lambda\lambda$1238.8,1242.8
appear unsaturated, but not
\ion{C}{4}~$\lambda\lambda$1548.2,1550.8 \citep[]{bibo84}.
Our grid of models shows
\ion{C}{4}~$\lambda\lambda$1548.2,1550.8 not saturated
when \Rstar$\ge$ 12\Rsun\space and \Teff$\gtrsim$42,000~K
(the actual temperature depends on \Rstar~ and \Mdot). 
Note that a large radius would be consistent
with the high luminosity of this object.

We studied the behaviour of the main lines
(\ion{C}{3}, \ion{C}{4}, \ion{N}{4}, \ion{N}{5},
\ion{O}{5}, \ion{O}{6} and \ion{P}{5})
in a wide range of temperatures, mass-loss rates and
\Lx values, ([37,000, 50,000]~K,
[2$\cdot10^{-7}$,1$\cdot10^{-6}$]\Myr, [-6.75,-8.0]) in search of
intervals where we can reproduce simultaneously all the
strongest features of the spectrum of HD~93250
and the unsaturated profile of
\ion{C}{4}~$\lambda\lambda$1548.2,1550.8.

We find that the lower the mass-loss rate,
the lower is the temperature where
models display unsaturated 
\ion{C}{4}~$\lambda\lambda$1548.2,1550.8.
In models with \Rstar=12\Rsun, independently of the value of shocks,
\ion{C}{4}~$\lambda\lambda$1548.2,1550.8 becomes
unsaturated  at \Teff=42,000/45,000/49,000~K for
\Mdot=$2\cdot10^{-7}/5\cdot10^{-7}/1\cdot10^{-6}$\Myr. 
However, the upper limit to \Teff~ set by
\ion{N}{4}~$\lambda\lambda$1718.0,1718.5 and
\ion{O}{6}~$\lambda\lambda$1031.9,1037.6
is lower (by 
$\sim$2,000/4,000/6,000~K and 4,000/7,000/9,000~K
respectively, for \Lx=-6.75) 
than required for the unsaturated
\ion{C}{4}~$\lambda\lambda$1548.2,1550.8.
If instead we decrease shocks
for an intermediate mass-loss rate of
\Mdot=$5\cdot10^{-7}$\Myr,
the temperature required to fit
\ion{O}{6}~$\lambda\lambda$1031.9,1037.6 
and \ion{N}{4}~$\lambda\lambda$1718.0,1718.5
increases by 1,000~K for every drop of
$\Delta\log$\Lx=-0.5 while it remains
constant for \ion{C}{4}~$\lambda\lambda$1548.2,1550.8.
Therefore in models with \Teff$\sim$42,000~K and
low \Mdot~ and \Lx~ this
gap would be reconciled. However, other prominent lines
(\ion{O}{5}~$\lambda$1371.0,
\ion{P}{5}~$\lambda\lambda$1118.0,1128.0 and
\ion{C}{3}~$\lambda$1176)
are not reproduced.
We conclude that we cannot fit the observed
\ion{C}{4}~$\lambda\lambda$1548.2,1550.8
consistently with the rest of the spectral lines.

We find the best solution to be $\log$\Lx=-7.25,
\mdot=5.5$\cdot10^{-7}$\Myr, \Teff=40,000~K, 
\Rstar= 12\Rsun\space and \logg=4.
With these parameters the model
reproduces \ion{O}{6}~$\lambda\lambda$1031.9,1037.6,
\ion{P}{5}~$\lambda\lambda$1118.0,1128.0,
\ion{C}{3}~$\lambda$1176, \ion{O}{5}~$\lambda$1371.0
and \ion{N}{4}~$\lambda\lambda$1718.0,1718.5 
(see Figures \ref{f_fit_all_fuse} and \ref{f_fit_all_iue}).
This model however, has saturated profiles of
 \ion{N}{5}~$\lambda\lambda$1238.8,1242.8 and
\ion{C}{4}~$\lambda\lambda$1548.2,1550.8, in
disagreement with the observed spectrum. 
Because \ion{C}{3}~$\lambda$1176
is well fit, a different carbon abundance would not improve
the solution.
As seen in Section \ref{s_fit_v},
\ion{N}{5}~$\lambda\lambda$1238.8,1242.8 is always
saturated in the range of parameters considered
for solar abundances, and models underabundant
in nitrogen are incompatible with
\ion{N}{4}~$\lambda\lambda$1718.0,1718.5.
The non-saturated absorptions of
\ion{C}{4}~$\lambda\lambda$1548.2,1550.8 and
\ion{N}{5}~$\lambda\lambda$1238.8,1242.8
may be due to an unresolved cooler companion.

\subsubsection{\it HD~64568}
\label{s_64568}

The only IUE spectrum of HD~64568 has
low dispersion. 
The stellar parameters are therefore constrained
primarily with lines in the FUSE range.
By consequence, the uncertainty of the derived
\Mdot~ and \Lx is higher than for the 
other objects of the sample, as we explain below.
This is not the case for \vinf~ (derived
from \ion{O}{6}~$\lambda\lambda$1031.9,1037.6)
or \logg (determined with
\ion{C}{3}~$\lambda$1176
and \ion{P}{5}~$\lambda\lambda$1118.0,1128.0,
but highly uncertain among the sample dwarf stars),
which are entirely determined from the high resolution
\textit{FUSE} spectrum.
The effective temperature is well constrained
with \ion{C}{3}~$\lambda$1176
and \ion{P}{5}~$\lambda\lambda$1118.0,1128.0.
These lines in HD~64568 are the weakest
in our sample indicating
higher \Teff~ (41,000~K), in accordance with its
earliest spectral type.

\ion{C}{4}~$\lambda\lambda$1548.2,1550.8 is unsaturated.
At the star's effective temperature, the models
(degraded to the [low] resolution of the observed spectrum)
only display unsaturated \ion{C}{4}~$\lambda\lambda$1548.2,1550.8
if \Mdot$\leq 2\cdot10^{-7}$\Myr,
\Rstar$>$12\Rsun\space and $\log$\Lx$>-7.0$.
\ion{C}{4}~$\lambda\lambda$1548.2,1550.8
decreases
with larger radii and, together with 
\ion{O}{6}~$\lambda\lambda$1031.9,1037.6,
helps us constrain this parameter.
We set the upper limit for the radius to 16\Rsun, when the models 
display
\ion{O}{6}~$\lambda\lambda$1031.9,1037.6 in excess
regardless of the other stellar parameters, and
the lower limit to 12\Rsun, when they cannot produce unsaturated
\ion{C}{4}~$\lambda\lambda$1548.2,1550.8.

The mass-loss rate and shocks were 
determined as explained in Section \ref{s_fit_v}.
Since most of the \mdot~ indicators 
(\ion{O}{5}~$\lambda$1371.0,
\ion{Si}{4}~$\lambda\lambda$1393.8,1402.8
and \ion{N}{4}~$\lambda\lambda$1718.0,1718)
are in the \textit{IUE} range,
the derived mass-loss rate of HD~64568 is
highly uncertain.
The derived \Lx, though determined from  
\ion{O}{6}~$\lambda\lambda$1031.9,1037.6
in the \textit{FUSE} range, is affected
by the uncertainty in \mdot, since
the strength of the doublet
also depends on mass-loss rate.

\subsection{\it The Supergiant Star, HD~190429A}
\label{s_190429} 
The spectrum of the O4 supergiant 
displays very strong lines which,
in comparison to the spectra of the
mid-O supergiant stars analysed in paper~I
(see Section \ref{teff_effects}), denote higher
temperature and mass-loss rate. 
In particular, the prominent
\ion{He}{2}~$\lambda$1640.0 emission line
indicates, according to our grid of models,
that the mass-loss rate
is higher than $7 \cdot 10^{-6}$\Myr\space.

In the range of temperatures and mass-loss rates
we considered
for HD~190429A
(35,000-40,000~K and 3.$\cdot 10^{-6}$
-1.5$\cdot 10^{-5}$\Myr,
as suggested by the star's spectral morphology),
\ion{C}{3}~$\lambda$1176, 
\ion{P}{5}~$\lambda\lambda$1118.0,1128.0 and
\ion{Si}{4}~$\lambda\lambda$1393.8,1402.8
increase
with increasing \Mdot~ and with decreasing \Teff~ and \Lx
(similarly to what we found in paper~I).
\ion{He}{2}~$\lambda$1640 has the same 
dependence on \Mdot~ and \Teff~
but does not vary appreciably with \Lx.
The effective temperature is constrained
from \ion{C}{3}~$\lambda$1176 and
\ion{Si}{4}~$\lambda\lambda$1393.8,1402.8.
For \Teff $\leq$ 36,000~K the models
with \Mdot $> 7 \cdot 10^{-6}$\Myr\space 
display a P~Cygni profile of \ion{C}{3}~$\lambda$1176,
not seen in the observed spectrum.
For \Teff $\geq$ 39,000~K, \ion{C}{3}~$\lambda$1176
and \ion{Si}{4}~$\lambda\lambda$1393.8,1402.8  are
absent even in models with mass-loss rate as high as 
\mdot $\sim$ 1.4 $\cdot$ 10$^{-5}$ \myr, thus providing
an upper limit to \Teff.
In this range of temperatures, the models reproduce
\ion{He}{2}~$\lambda$1640.0 for
\mdot $\geq$ 1.0 $\cdot$ 10$^{-5}$ \myr~
(lower limit of mass-loss rate).
Models with \mdot $\geq$ 1.4 $\cdot$ 10$^{-5}$ \myr~
(upper limit) display excessive
\ion{P}{5}~$\lambda\lambda$1118.0,1128.0 profiles.
The shocks in the wind were determined by fitting
\ion{O}{6}~$\lambda\lambda$1031.9,1037.6
and
\ion{O}{5}~$\lambda$1371.0 (both oxygen lines increase
with increasing shocks) and 
\ion{Si}{4}~$\lambda\lambda$1393.8,1402.8.
In addition, different values of \Lx
change the shape of the
\ion{P}{5}~$\lambda\lambda$1118.0,1128.0 absorptions.

In this regime of \Teff~ and \Mdot,
the models with \Rstar$\ge$ 23 \Rsun~ produce 
unsaturated \ion{C}{3}~$\lambda$1176 
thus setting the lower limit of \Rstar.
The upper limit is 27\Rsun~ because models with larger radius
produce \ion{O}{6}~$\lambda\lambda$1031.9,1037.6
stronger than observed
regardless of the other stellar parmeters.
Other lines sensitive to the radius, which helped us to further
contrain its value,
are \ion{He}{2}~$\lambda$1640.0 and 
\ion{Si}{4}~$\lambda\lambda$1393.8,1402.8
(which decrease with increasing radius)
and \ion{O}{5}~$\lambda$1371.0 (favoured by larger radii); 
the profile of \ion{P}{5}~$\lambda\lambda$1118.0,1128.0
+\ion{Si}{4}~$\lambda\lambda$1122.5,1128.3,1128.4 
also changes with radius.

Contrary to what we find for the dwarf stars, almost all the
spectral lines are sensitive to gravity, and
we can use them to derive \logg. Higher \logg corresponds
to weaker \ion{O}{6}~$\lambda\lambda$1031.9,1037.6,
\ion{O}{5}~$\lambda$1371.0
and \ion{He}{2}~$\lambda$1640.0, while
\ion{Si}{4}~$\lambda\lambda$1393.8,1402.8 increases and
the aborption of
\ion{P}{5}~$\lambda\lambda$1118.0,1128.0 grows
deeper. The carbon lines are also affected by an increase of
\logg: \ion{C}{4}~$\lambda\lambda$1548.2,1550.8
decreases and the absorption of
\ion{C}{4}~$\lambda$1169+\ion{C}{3}~$\lambda$1176
increases whereas its emission diminishes.

\begin{figure}
\epsscale{0.75} \rotatebox{180}{\plotone{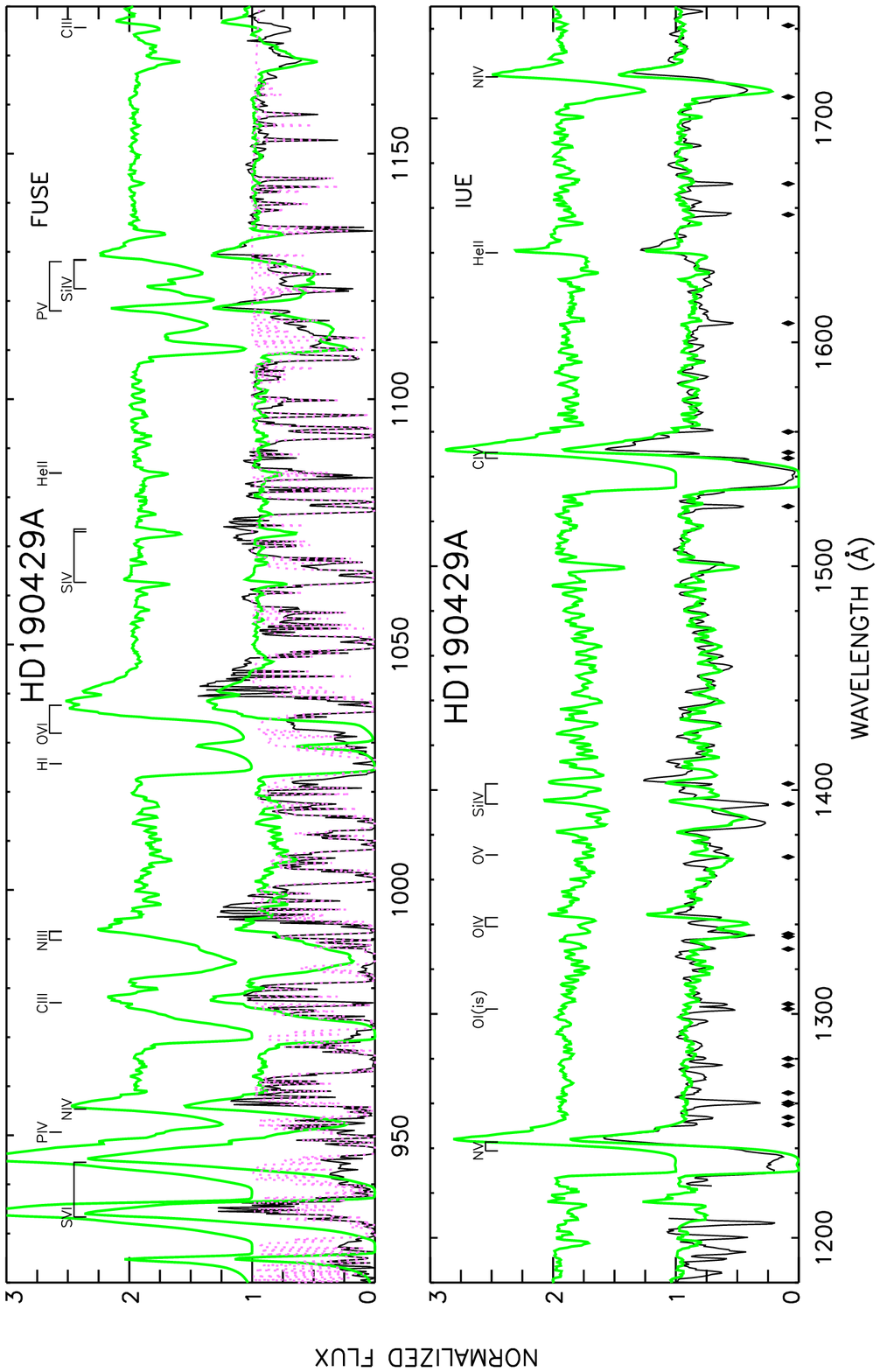}}
\raisebox{-9in}{\rotatebox{90}{ \parbox[t]{9in}{
\caption[f7.eps]{
Best fit models of
HD~190429A (top: \textit{FUSE} spectra, 
bottom: \textit{IUE} spectra). Two models
(green/grey) are shown, 
both with \Teff =37,500~K, log = 3.4, R= 25\Rsun, \mdot = 1.2 $\cdot$
10$^{-5}$ \myr, $\log$\Lx = -6.5  , \Lbol=6.05 and \Vinf = 2100
\kms.
The model superimposed to the spectrum, with solar abundances, provides
a good fit of most wind features in both \textit{FUSE}
and \textit{IUE} range.
The top model has modified CNO abundances
($\epsilon _{He} = 5 \; \epsilon _{He,\odot}$,
$\epsilon _C = 0.5 \; \epsilon _{C,\odot}$,
$\epsilon _N = 2.0 \; \epsilon _{N,\odot}$,
$\epsilon _O = 0.1 \; \epsilon _{O,\odot}$)
and fits better
\ion{C}{4}~$\lambda$1169+\ion{C}{3}~$\lambda$1176
and \ion{O}{6}~$\lambda\lambda$1031.9,1037.6,
but less well
\ion{P}{5}~$\lambda\lambda$1118.0,1128.0
and
\ion{Si}{4}~$\lambda\lambda$1393.8,1402.8.
\label{f_hd190429_col}
}}}}
\end{figure}

We find the best fit at \Teff =37,500~K, \logg = 3.4, \Rstar= 25\Rsun,
\mdot = 1.2 $\cdot$ 10$^{-5}$ \myr, $\log$\Lx = -6.5  , \Lbol=6.05 and
\Vinf = 2100 \kms. 
This model reproduces most of the wind 
features (see Figure \ref{f_hd190429_col})
including the lines of \ion{O}{4}, \ion{O}{6},
\ion{P}{5}+\ion{Si}{4}, \ion{C}{4},
\ion{Si}{4}, \ion{He}{2} and \ion{N}{4}.
Minor discrepancies remain in the fit to some of these lines,
which can be attributed to the contribution of
the secondary to the flux of the system.
Even though the lines used for the analysis
come primarily from the O4 star, they may have a residual flux 
in the absorption trough from the O9.5~II companion. 
The comparison of spectra from the IUE atlas of stars \citep[]{IUEatlas}
of O4~I stars (main component) and
O9.5~II stars (companion) 
provides an upper limit for the contribution of
the secondary.
An O9.5 bright giant displays weak 
\ion{N}{5}~$\lambda\lambda$1238.8,1242.8 
while the doublet has typically a saturated absorption in an O4~I star.
From the depth of \ion{N}{5}~$\lambda\lambda$1238.8,1242.8
in the IUE spectrum of the system we estimated that the
companion may amount to $\lesssim$ 15\% of the total.
The O9.5~II star has strong 
\ion{Si}{4}~$\lambda\lambda$1393.8,1402.8; this
would explain why 
\ion{Si}{4}~$\lambda\lambda$1393.8,1402.8
is weaker in our models than observed.
\ion{C}{4}~$\lambda\lambda$1548.2,1550.8
is fully developed but not saturated in the O9.5~II star,
which would explain the excess emission and absorption
of the model. The spectrum of an O9.5~II has
weak photospheric \ion{N}{4}~$\lambda\lambda$1718.0,1718.5,
and this makes the P~Cygni profile
from the O4 If$^+$ star
look less developed (relative to the total continuum),
thus this line appears slightly in excess in the model.

In the FUSE range, according to the Copernicus atlas of OB stars
of \citet[]{W96}, an O9.5~II star has photospheric
\ion{S}{4}~$\lambda\lambda$1062.7,1073.0,1073.5,
and \ion{P}{5}~$\lambda\lambda$1118.0,1128.0
+\ion{Si}{4}~$\lambda\lambda$1122.5,1128.3,1128.4 lines,
which will increase the depth of the O4 If$^+$ wind lines.
\ion{O}{6}~$\lambda\lambda$1031.9,1037.6
and \ion{C}{4}~$\lambda$1169+\ion{C}{3}~$\lambda$1176
have unsaturated P~Cygni profiles in O9.5~II stars,
possibly causing the mismatch of the best fit model
to these lines.

However, there is another explanation for the discrepancies.
It is possible that HD~190429A has non solar
surface abundances.
In Figure \ref{f_hd190429_col}
we present a model with modified CNO abundances 
($\epsilon _{He} = 5 \; \epsilon _{He,\odot}$,
$\epsilon _C = 0.5 \; \epsilon _{C,\odot}$,
$\epsilon _N = 2.0 \; \epsilon _{N,\odot}$,
$\epsilon _O = 0.1 \; \epsilon _{O,\odot}$),
which would be compatible with a
scenario where HD~190429A is evolving towards the Wolf-Rayet phase,
as suggested by previous authors (see Section \ref{s_stars}).
The modified CNO abundances improve the fit to
\ion{C}{4}~$\lambda$1169+\ion{C}{3}~$\lambda$1176
and the oxygen lines
(\ion{O}{6}~$\lambda\lambda$1031.9,1037.6,  
\ion{O}{4}~$\lambda\lambda$1339.0,1343.0
and \ion{O}{5}~$\lambda$1371.0) but
degrades the fit to
\ion{P}{5}~$\lambda\lambda$1118.0,1128.0 and
\ion{Si}{4}~$\lambda\lambda$1393.8,1402.8.

There is one feature,
\ion{S}{4}~$\lambda\lambda$1062.7,1073.0,1073.5,
definitely produced in the primary star spectrum,
that the \textit{WM-basic} models do not reproduce.
We encountered the same problem when we modeled
the spectra of the mid-O type supergiant stars in
paper~I. 
The \ion{S}{4}~$\lambda\lambda$1062.7,1073.0,1073.5
problem is probably due to the atomic
data used in \textit{WM-basic} and prompted a comparison
with models computed with
the  \citet[]{cmfgen98}
\textit{CMFGEN} code. The synthetic spectra generated
with both codes for an equivalent set of stellar
parameters display very similar
photospheric and wind features with the exception
of \ion{S}{4}~$\lambda\lambda$1062.7,1073.0,1073.5 
(\citet[]{BGH03}; Bianchi, Garcia, \& Herald, in preparation).
\textit{WM-basic} and \textit{CMFGEN} predict
very different ionization fractions for sulphur.

\section{SUMMARY AND DISCUSSION}
\label{s_conclusion}

We have performed a detailed spectroscopic analysis
of seven Galactic early-O type stars and
derived their stellar parameters.
We found effective temperatures ranging
from 41,000~K to 39,000~K for the O3-4 dwarf stars and
37,500~K for the O4 supergiant.
The X-ray ionization due to shocks in the wind 
has been constrained by our modeling,
primarily from the
\ion{O}{6}~$\lambda\lambda$1031.9,1037.6
resonance doublet,
accessible with the \textit{FUSE} telescope.
In the range
of \Teff, \Mdot~ and \Rstar~ that we derived for
the dwarf stars, the spectral
morphology hardly changes with  gravity and
the derived value of \logg is less well constrained
than other parameters, but consistent
with previous estimates.

All the sample stars display unsaturated
\ion{N}{5}~$\lambda\lambda$1238.8,1242.8
profiles that we cannot fit
consistently with the rest of the spectrum.
In one case at least (HD~96715), the unsaturated profile is 
due to poor background subtraction in the IUE data
(as shown by an \textit{ORFEUS} spectrum),
and this may apply to the other stars as well.
In the case of HD~190429A, HD~93205 and HD~168076,
the unsaturated profile may also be due to
flux from the cooler companion star,
whose contribution we estimate to be
$\leq$ 15\% of the observed flux.
A comparison of \textit{WM-basic} and \textit{CMFGEN}
models calculated with parameters suitable for 
the early-O type stars is under way
(\citet[]{BGH03}; Bianchi, Garcia, \& Herald, in preparation),
to investigate the effect of a different treatement
of the shocks on the
\ion{N}{5}~$\lambda\lambda$1238.8,1242.8
doublet.

We compile previous determinations of
the stellar parameters for our program stars
from the literature
in Table \ref{t_prev}. 
Our effective temperatures are consistently lower
than those derived in other works, with the only exception
of the Str\"omgren and H$\beta$ photometric studies  by \citet[]{KG93}
and \citet[]{SJ95}.
\citet[]{KG93} derived temperatures for the Carina
stars of our sample lower than ours by
$\sim$4,000~K on average.
\citet[]{SJ95} assigned a temperature
of 28,200~K to HD~190429 (without resolving
the system),
closer to what we would expect for the secondary
than to what we found for HD~190429A (37,500~K).
\citet[]{M75} performed Str\"omgren
and H$\beta$ photometry of HD~168076 and
derived a temperature higher than
ours by 16,000~K.
The calibration of \citet[]{C75}, based on
the comparison of the equivalent widths of
hydrogen and helium lines with plane-parallel
hydrostatic non-LTE model predictions, was used by
\citet[]{CW76} and \citet[]{C77} for several
objects of our sample, yielding
temperatures higher than our results by
$>$ 10,000~K.
Temperatures obtained from fitting
Balmer lines and optical \ion{He}{1} and \ion{He}{2} lines
with plane-parallel non-LTE hydrostatic models
are $\sim$  10\% \citep[]{Sal83} to $\sim$20\%
\citep[]{Kal92,puls96} higher than ours.
\citet[]{WMBAS} compared the \textit{IUE} spectrum
of HD~93250 to \textit{WM-basic} models 
of \Teff=50,000~K (approximately the same as 
derived by \citet[]{puls96}), \logg=4.0 and \Rstar=12\Rsun,
with different mass-loss rates around
\Mdot=5.6$\cdot 10^{-6}$\Myr. 
Their synthetic spectra did not include shocks
effects and either produce
\ion{O}{5}~$\lambda$1371.0 in excess or cannot reproduce the
\ion{N}{4}~$\lambda\lambda$1718.0,1718.5 line.
In our work we used those lines, together with
\ion{C}{4}~$\lambda$1169+\ion{C}{3}~$\lambda$1176
and \ion{P}{5}~$\lambda\lambda$1118.0,1128.0,
to set the upper limit of the effective temperature
of the dwarf stars
to $\sim$42,000~K (see Section \ref{s_models})
and we obtained a consistent fit of all spectral features
by including the effects of shocks in the calculations.

\begin{figure}
\epsscale{0.75}
\plotone{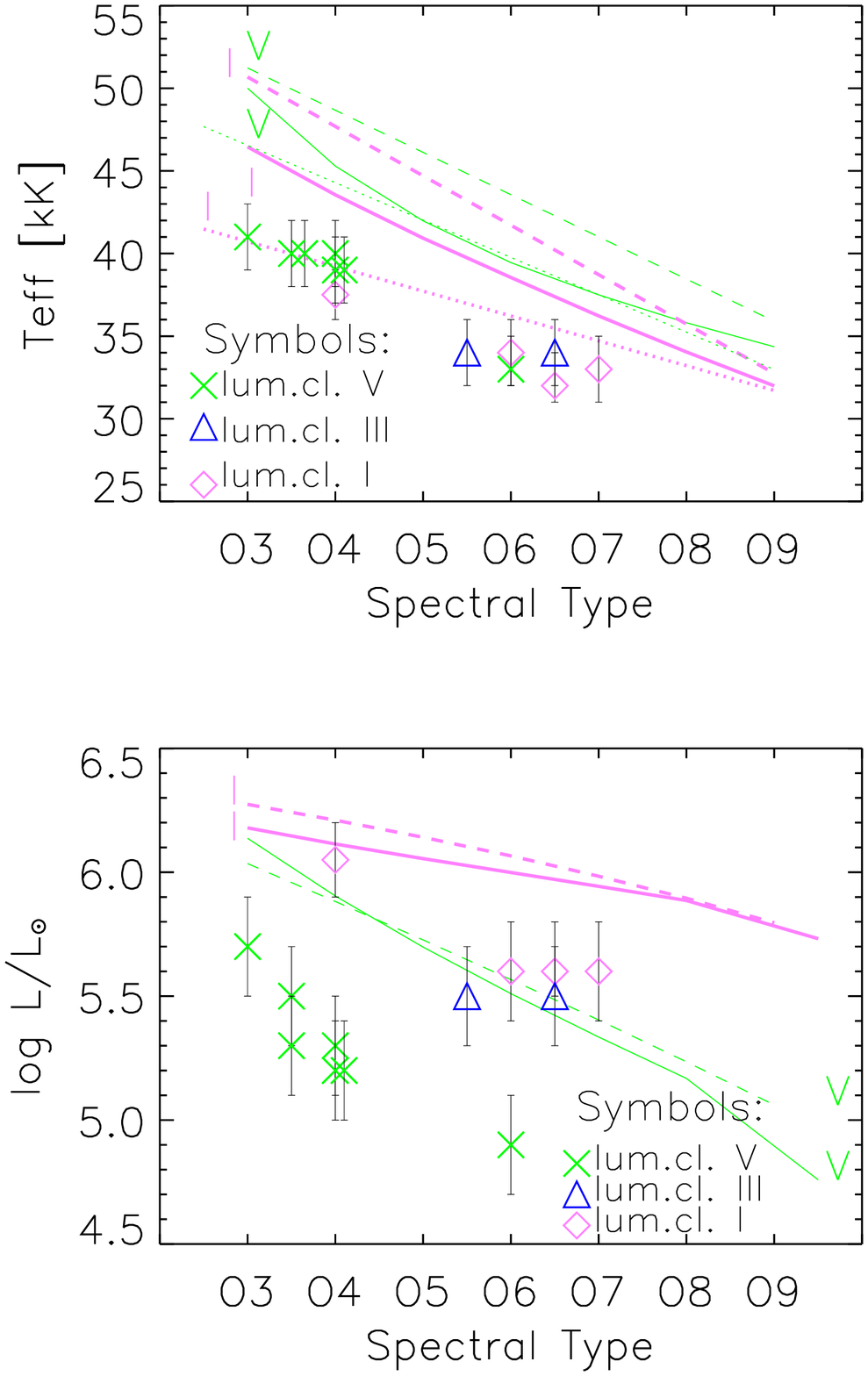}
\caption[f8.eps]{
\scriptsize{
\Teff~ (top plot) and \Lbol~ (bottom plot)
obtained for the objects analysed in this paper and in \citet[]{BG02}
are compared to the empirical calibrations of
\citet[]{vacca} (dashed line), \citet[]{dejag} (solid line) and
\citet[]{MPRM03} (dotted line). The thick lines
represent the curves for supergiants and the thin lines represent dwarf stars;
the symbols for individual objects indicate luminosity class
and are explained in the figures. 
Our temperatures and luminosities are lower
than these previous calibrations.
}
\label{tefflum}}

\end{figure}

In Figure \ref{tefflum} we compare the effective temperatures
and luminosities obtained for the total sample
of O3-O7 type stars
analysed in this work and in paper~I
with previous empirical calibrations
(\citet[]{vacca}, \citet[]{dejag} and \citet[]{MPRM03}).
Our \Teff~ values for O3-O6 dwarf stars are 
lower than these calibrations, the differences ranging between
9,000-10,000~K, 6,000-9,000~K and 4,000-7,000~K respectively.
The discrepancy is smaller for the O4-O7 supergiants
(6,000-10,000~K, 4,000-6,000~K and 2,000-3,500~K, respectively).
The derived luminosities 
for both dwarf stars and supergiants
are also lower, by an average of 7\% and 10\% respectively,
than both the calibrations of
\citet[]{dejag} and \citet[]{vacca}.
For the O4~If$^+$ supergiant the luminosity is
also lower, but the discrepancy is within the error bars.

The lower temperatures from our analysis
are due to two main improvements.
On the one hand, the use of \textit{FUSE} data
enables the assessment of X-rays from shocks and thus a correct
(consistent) derivation of the wind ionization.
On the other hand, the inclusion of line-blanketing
effects in the analysis
yields lower effective temperatures than
pure hydrogen and helium model analyses.
Optical spectra analyses of massive stars with
spherical, hydrodynamic,
wind-blanketed, non-LTE synthetic spectra 
also revised the temperature
scale downwards \citep[]{MSH02,HPN02,RPH03}.
Our derived \Teff~ values are still 
lower.
Temperatures 
of O3~V and O4~V stars, according to the scale
of \citet[]{MSH02} (based on calculations with the CMFGEN code)
are $\simeq$47,500~K and $\simeq$44,500~K,
whereas we have obtained 41,000~K and 40,000-39,000~K.
From the work of \citet[]{HPN02}, who fit Balmer, \ion{He}{1}
and \ion{He}{2} optical lines with FASTWIND \citep[]{FASTWIND}, we can interpolate
a temperature of 41,250~K for an O4~If$^+$, i.e.
$\sim$4,000~K higher than we derived for HD~190429A.
\citet[]{RPH03}, proceeding similarly, obtained
46,000~K for HD~93250 (in contrast with our result, 40,000~K)
and 41,000~K for HDE~303308 (in agreement within the
error bars with our value, 40,000~K).
The remaining disagreement between these 
results and ours may originate in the fact
that the analyses are based on optical and UV data respectively.
We are planning to analyse
optical spectra of the sample stars
to check the consistency of our results from the UV and Far-UV.

The spectral lines in the FUSE and IUE ranges 
constrain \Mdot~ as well.
The mass-loss rate we derived for HD~190429A 
(1.2$\cdot 10^{-5}$\Myr)
agrees with the values derived from H$\alpha$ by
\citet[]{C77} ($1.1\cdot10^{-5}$\Myr),
\citet[]{L88} (1.26$\cdot 10^{-5}$\Myr), \citet[]{Sal92} (8.6$\cdot 10^{-6}$\Myr)
and \citet[]{MPRM03} ($1.42\cdot10^{-5}$\Myr).
Studies of IR and radio
spectra yield lower \Mdot\space: $4.6\cdot10^{-6}$\Myr~ \citep[radio]{Sal98},
$<$0.35$\cdot 10^{-6}$\Myr \citep[IR]{PFG83}.
For the dwarf stars the
mass-loss rates derived in the literature are at least one order of
magnitude higher than ours. For HD~93250,
the exceptionally high value of \Mdot=$4.1\cdot10^{-5}$ \citep[]{Lal95}
was determined from radio fluxes at 8.54 GHz,
but the data may be contaminated by non-thermal emission;
for this object, there is also one measurement lower than ours by one order of magnitude.
In view of the systematic differences for dwarf stars,
we used the ``recipe'' of \citet[]{Vink},
based on the radiation pressure driven wind theory,
to predict the mass-loss rates of the sample stars
from their derived photospheric parameters. The predictions are
presented in Table \ref{t_models}.
The mass-loss rate derived for HD~190429A agrees
with the prediction. For dwarf stars our values are
lower than predicted by a factor of $\sim$2
($\sim$4 for HD~64568).

In general, the stellar masses we have obtained are
lower than values determined in
previous works, as shown in Table \ref{t_prev}.
In the case of HD~93205, the mass had been
derived from the orbital parameters of the binary
system:
\Mstar$\ge$ 31.5 \citep[]{Mal01},
\Mstar$\ge$ 32.6 \citep[]{SL93},
\Mstar$\ge$ 37 \citep[]{Cal99}. 
For this star we derive 30$\pm10$\Msun,
lower than previous determinations but
compatible within the errors.
The primary mass yields a mass of the secondary
of 13$\pm5$\Msun~ for
a mass ratio of q=0.423 \citep[]{Mal01}, a mass
significantly lower than derived for other O8~V stars
in eclipsing systems like, for example, DN~Cas, for which
\citet[]{dncas} give 24\Msun.
However, it must be emphasized here that
the uncertainty in \logg and \Rstar~
(see Section \ref{s_fit_v}) makes
the derived value of the stellar masses much
more uncertain than other parameters
(like \Teff~ and \Mdot).
Hopefully the mass determination will be refined with a
consistent analysis of optical spectra, planned
as a future work.

\begin{figure}
\epsscale{1.}
\plotone{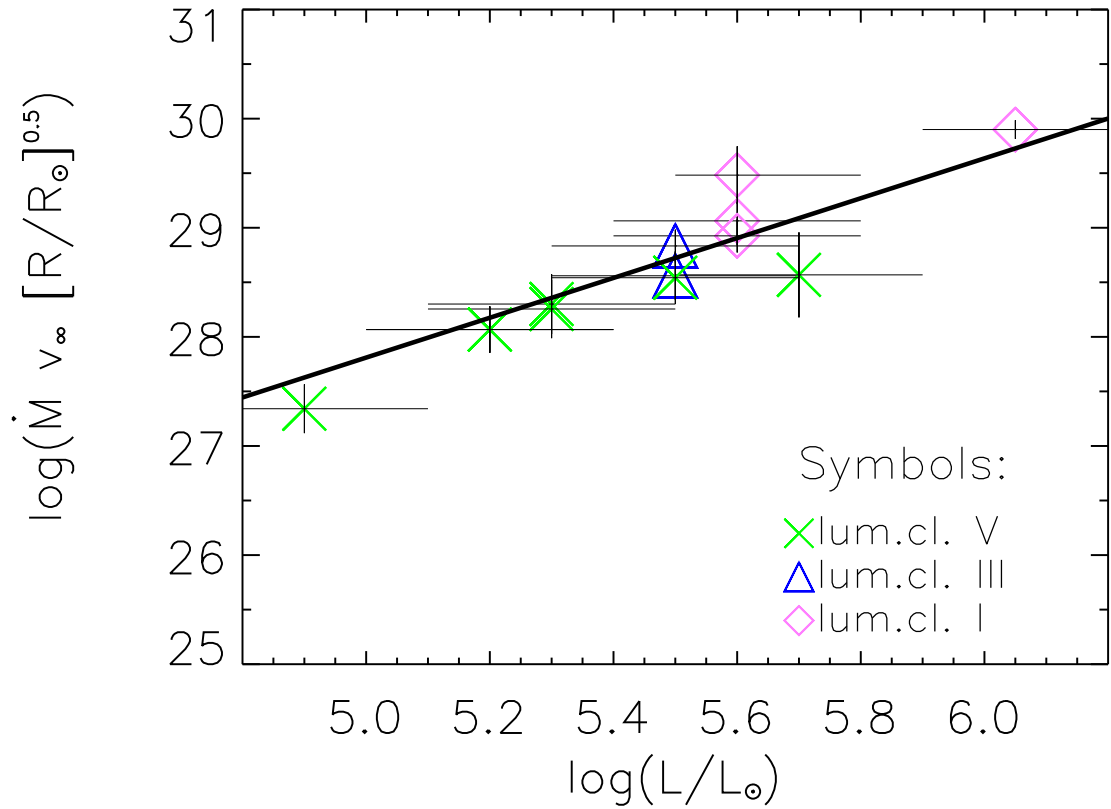}
\caption[f9.eps]{
Modified wind-momentum \textit{versus} luminosity
for the objects of the present sample 
plus those analysed in paper~I.
The product \Mdot$\cdot$\Vinf~ is in cgs-units.
Symbols represent individual objects of different
luminosity classes and are explained in the plot.
The solid line represents the
theoretical WLR of \citet[]{Vink}.
\label{windmom}}

\end{figure}

Our results improve the empirical
calibration of the Wind momentum-Luminosity Relation (WLR). 
The WLR relates the so-called modified wind momentum
$D_{mom}=$ \Mdot \Vinf $(R_{\ast}/R_{\odot})^{0.5}$
to the luminosity of the star as
$D_{mom} \propto L^x$ \citep[]{Kal95}.
In the past few years a big effort
has been made to calibrate the WLR
\citep[]{puls96,KP00,HPN02,RPH03,MPRM03} for its promising
application to distance determinations. 
In Figure \ref{windmom} we plot the WLR for
the total sample of stars studied in this paper and in
paper~I. Our points are mostly consistent with
the theoretical relation of \citet[]{Vink}
within the error bars.

We are planning to extend our modeling of
Far-UV and UV spectra to a larger sample of
O type stars and to the optical range.
The lower \Teff~ values (compared
with previous results/calibrations) have
great implications for understanding
the ionisation of $HII$ regions,
since cooler stars would produce
less photons energetic enough to ionize hydrogen
($\lambda<$ 912\AA).
For instance, the ionizing flux
from a star of \Teff=40,000~K amounts to only
30\% of the flux from a \Teff=50,000~K star
with the same radius.

\acknowledgments
We are indebted to Adi Pauldrach for assisting us 
with the use of the \textit{WM-basic} code.
This work is based on data obtained 
by the NASA-CNES-CSA {\it FUSE} mission operated by
 the Johns Hopkins University.
The data from the \textit{IUE} satellite presented
in this paper were obtained from both the INES archive and
the Multimission Archive at the Space Telescope Science
Institute (MAST).
Financial support has been provided in part  by 
NASA grants NAS5-32985 and NRA-99-01-LTSA-029.


\begin{deluxetable}{lcccc}
\tablecaption{Datasets used in this work \label{t_id} }
\tablewidth{0pt}  
\tablecolumns{5}
\tablehead{
\colhead{Star} &
\colhead{\textit{FUSE}} &
\colhead{\textit{IUE}} &
\colhead{\textit{HST-STIS}} &
\colhead{\textit{ORFEUS}}
}
\startdata
HD~190429A & P1028401    & SWP38994  &       &       \\
HD~64568  & P1221103    & SWP29337  &       &       \\
HD~93250  & P1023801    & SWP22106  &       &       \\ 
HD~93205  & P1023601    & SWP14747  & O4QX01010 &       \\
HDE~303308& P1221602    & SWP07024  & O4QX04010     &       \\
HD~96715  & P1024301    & SWP43980,SWP21999  &       & tues5233\_4   \\
HD~168076 & P1162201    & SWP28277  &       &       \\
\enddata

\end{deluxetable}


\pagestyle{empty}
\renewcommand{\arraystretch}{.6}
\tabletypesize{\scriptsize}
{
\rotate
\begin{deluxetable}{lllcccccc}
\tablecaption{The program stars \label{t_stars} }
\tabletypesize{\scriptsize}
\tablewidth{0pt}  
\tablecolumns{9}
\tablehead{
\colhead{Star} &
\colhead{Sp. Type  \tablenotemark{a}} &
\colhead{Sp. Type Ref.} &
\colhead{ $V$ } &
\colhead{ \bv } &
\colhead{ E(\bv) \tablenotemark{b}} &
\colhead{ $\log{N_{HI}}$ \tablenotemark{c}} &
\colhead{ $\log{N_{H_{2}}}$ \tablenotemark{d}} &
\colhead{ Distance (Kpc) } 
\\
\colhead{ } &
\colhead{ } &
\colhead{ } &
\colhead{ } &
\colhead{ } &
\colhead{ } &
\colhead{ [$cm^{-2}$] } &
\colhead{ [$cm^{-2}$] } &
\colhead{ [Kpc] } 
}
\startdata
HD~190429A&  \textbf{O4 If$^+$}; O4 f; O4 I & (1,2,3;4;5)
     &7.07(6)  & 0.09 (6)   & 0.41 & 21.29& 20.31& 2.3 (7)  \\
     & O5~If+O9.5~III+B1~IIIs& (8)    &    &    &  &   &  \\
     & O4f+O9.5III ; O4If$^+$+O9.5 II& (9,10;11)  &     &     &    &  &   &  \\
     & O5f+B0III ; O5f+O9.5Ibp & (12;13)    &   &    &  &   &     &  \\

HD~64568  & \textbf{O3 V((f$^*$))}; O4 V ((f)); O5 V & (14,1;15;16)
            & 9.39 (17) & 0.11 (17)   & 0.44 
                            & 21.32& 20.34 &5.5 (18)\\
     & O5; O5/6 & (19,20,21,22;23)  &       &   &    &  &   &      \\
HD~93250  & \textbf{O3.5 V((f$^+$))}; O3~V((f)); O3:V((f)); O3~V(f)& (1;2,14,21,24;25;26)  &7.37(27)&0.17(27)& 0.50 &21.38
                &19.92/19.94\tablenotemark{e} (28)& 2.5 (25) \\
     & O3~I; O3~III(f);O3 ; O5 & (5;29;4;13) &      &   &    &  &   &      \\
HD~93205  & \textbf{O3.5 V((f$^+$))}; O3~V; O3& (1;14,21,2,25,24,26;4)
        & 7.75 (30) & 0.05 (30) & 0.38  & 21.33 (31)& 19.52 &  2.5 (25) \\
     & O3~V+O8~V&(32,33) &         &   &    &  &   &      \\
HDE~303308& \textbf{O4 V((f$^+$))};O4~V((f))& (1;29) & 8.15(27) & 0.12(27) & 0.45 
                            &21.45 (31) & 19.99/19.81\tablenotemark{e} (28) & 2.5 (25) \\
     & O3~V; O3~V((f)); O3~V(f);O3 & (24,2;3,14,21,25;26;4)        &   &   &    &  &   &      \\
HD~96715  & \textbf{O4 V((f))}; O4~V; O5: & (3,21;2,15; 26) & 8.27 (34) & 0.10 (34) & 0.43 
                            & 21.31& 20.33 & 2.9 (3) \\
HD~168076 &\textbf{O4~V((f))}; O4~V((f$^{+}$)); O5~V((f$^{*}$)) & (3;35;36)
        & 8.18 (36) & 0.43 (36) & 0.75 (36) & 19.67/21.65\tablenotemark{f} (37) & 20.32/20.43\tablenotemark{f} (37)
                    & 2.0 (36) \\
     & O4 ((f)); O4f ;O5 ; O4~III(f) &  (10,4;9;13,23;29) &    &   &    &  &   &      \\
\enddata

\tablenotetext{a}{Adopted types in bold face; other values
from the literature are compiled. For multiple systems we also list here the
spectral type of the companion (if known).}
\tablenotetext{b}{If no reference provided,
calculated from \bv~ (this table) and $(B-V)_0$ (from \citet[]{Massey98}).
}
\tablenotetext{c}{If no reference provided,
$N_{HI}$ is derived from the relation of
\citet[]{BO78}, using E(\bv) from this table.}
\tablenotetext{d}{If no reference provided,
$N_{H_2}=0.5(N_H-N_{HI})$ and 
$N_H/E(\bv) = 5.8 \cdot 10^{21} \;
atoms \; cm^{-2} mag^{-1}$ \citep[]{BO78}, using $N_{HI}$ and E(\bv) from this table.
}
\tablenotetext{e}{Column densities of the foreground cloud
(first value) and the Carina nebula (second value).}
\tablenotetext{f}{Column densities of the
traslucent component of the hydrogen cloud
(first value) and of the diffuse component (second value).}

\tablerefs{ (1) \citet[]{W02a}, (2) \citet[]{W72}, (3) \citet[]{W73}, 
 (4) \citet[]{C77},(5) \citet[]{Petal96}, (6) \citet[]{Bal00}, (7) \citet[]{morg53},
 (8) \citet[]{A86}, (9) \citet[]{CA71}, (10) \citet[]{CL74},
(11) \citet[]{WH00}, (12) \citet[]{guetter68}, (13) \citet[]{morg55},
(14) \citet[]{W82}, 
(15) \citet[]{garr77}, (16) \citet[]{P-J81}, (17) \citet[]{H72}, 
(18) \citet[]{KH00}, (19) \citet[]{mac76}, (20) \citet[]{cram}, (21) \citet[]{cruz74},
(22) \citet[]{lo65}, (23) \citet[]{mss}, (24) \citet[]{W71a},
(25) \citet[]{LM82}, (26) \citet[]{TA74}, (27) \citet[]{F82}, 
(28) \citet[]{Lal00}, (29) \citet[]{mathys88}, (30) \citet[]{F73},
(31) \citet[]{DS94}, (32) \citet[]{CW76}, (33) \citet[]{Mal01}, (34) \citet[]{Sch83}, 
(35) \citet[]{BMN99}, (36) \citet[]{ngc6611}, (37) \citet[]{rachford}
}

\end{deluxetable}
}



{\rotate
\begin{deluxetable}{llccccccccc}
\footnotesize
\tablecaption{Stellar parameters from model fitting of Far-UV
and UV spectra.\label{t_models} }
\tabletypesize{\footnotesize}
\tablewidth{0pt}  
\tablecolumns{11}
\tablehead{
\colhead{Star} &
\colhead{Sp type } &
\colhead{\Teff} &
\colhead{log~g} &
\colhead{R/R$_{\odot}$} &
\colhead{M/M$_{\odot}$} &
\colhead{log(L$_{bol}$/L$_{\odot}$)} &
\colhead{\Vinf } &
\colhead{\mdot \tablenotemark{a}} &
\colhead{\mdot \tablenotemark{b}} &
\colhead{\Lx}\\
\colhead{ } &
\colhead{ } &
\colhead{[kK] } &
\colhead{ } &
\colhead{ } &
\colhead{ } &
\colhead{ }  &
\colhead{[\kms ] } &
\colhead{[\Myr ]} &
\colhead{[\Myr ]} &
\colhead{ }
}
\startdata

HD~190429A    &O4 If$^+$  
        &37.5$\pm$1.5   & 3.4$\pm$.1    &25$\pm$2   & 57  & 6.05$\pm$.15    & 2100$\pm$200  & 
  1.2$\pm$.2 $\cdot$ 10$^{-5}$ & 9.2 $\cdot$ 10$^{-6}$ &
 -6.5$\pm$.5\\
HD~64568    &O3 V((f$^*$))  
        & 41$\pm$2  & 3.9$\pm$.2    &14$\pm$2   & 57  & 5.7$\pm$.2  & 2800$\pm$200  & 
  5.6$\pm$5. $\cdot$ 10$^{-7}$ & 2.3 $\cdot$ 10$^{-6}$  &
 -7.5$\pm$.75\\
HD~93250    &O3.5 V((f$^+$))
        & 40$\pm$2  & 4.0$\pm$.2    &12$\pm$2   & 53 & 5.5$\pm.2$   & 2900$\pm$200  & 
  5.5$\pm$3. $\cdot$ 10$^{-7}$ & 9.9 $\cdot$ 10$^{-7}$  &
 -7.25$\pm$.25\\
HD~93205    &O3.5 V((f$^+$))
        & 40$\pm$2  & 4.0$\pm$.2    &9$\pm$3   & 30 & 5.3$\pm.2$   & 3200$\pm$200  & 
  3.3$\pm$2. $\cdot$ 10$^{-7}$ & 5.1 $\cdot$ 10$^{-7}$  &
 -7.0$\pm$.5\\
HDE~303308  &O4 V((f$^+$))  
        & 40$\pm$2  & 4.0$\pm$.2   &9$\pm$3    & 30 & 5.3$\pm.2$   & 2800$\pm$200  & 
  3.4$\pm$2. $\cdot$ 10$^{-7}$ & 6.0 $\cdot$ 10$^{-7}$  &
 -7.0$\pm$.5\\
HD~96715    &O4 V((f))  
        & 39$\pm$2  & 4.0$\pm$.2   &9$\pm$3    & 30 & 5.2$\pm.2$   & 2800$\pm$200  & 
  2.2$\pm$1. $\cdot$ 10$^{-7}$ & 4.0 $\cdot$ 10$^{-7}$  &
 -7.25$\pm$.25\\
HD~168076   &O4~V((f))  
        & 39$\pm$2  & 4.0$\pm$.2    & 9$\pm$3   & 30 & 5.2$\pm.2$   & 2800$\pm$200  & 
  2.2$\pm$1. $\cdot$ 10$^{-7}$ & 4.0 $\cdot$ 10$^{-7}$  &
 -7.0$\pm$.25\\

\enddata
\tablenotetext{a}{ This paper; values from spectral modeling}
\tablenotetext{b}{ Mass-loss rate value predicted with the mass-loss ``recipe'' 
of \citet[]{Vink},
using the stellar parameters from our model fitting (this table)
} 
\end{deluxetable}
}


\pagestyle{empty}
\renewcommand{\arraystretch}{.6}
\tabletypesize{\scriptsize}
{
\rotate
\begin{deluxetable}{lcccccccccccc}
\tablecaption{The parameters of the program stars 
previously derived in the literature\label{t_prev} }
\tabletypesize{\tiny}
\tablewidth{0pt}  
\tablecolumns{13}
\tablehead{
\colhead{Star} &
\colhead{\Teff} &
\colhead{log g} &
\colhead{log(L/\lsun)} &
\colhead{Ref.} &
\colhead{R/\Rsun} &
\colhead{M/\Msun } &
\colhead{Ref.} &
\colhead{\Mdot \tablenotemark{a}} &
\colhead{\vinf} &
\colhead{Ref.} &
\colhead{\Lx} &
\colhead{Ref.} \\
\colhead{ } &
\colhead{[kK]} &
\colhead{ } &
\colhead{ } &
\colhead{ } &
\colhead{ } &
\colhead{ } &
\colhead{ } &
\colhead{[\Myr]} &
\colhead{[\kms ]} &
\colhead{ } &
\colhead{ } &
\colhead{ }
}
\startdata
HD~190429A  & 37.5 & 3.4 & 6.05 &(1)& 25 & 57 &(1)& 1.2$\cdot10^{-5}$ & 2100 &(1)& -6.5 &(1) \\
        & 28.2 & 3.53 &   & (2) &   &   & &
    $4.6\cdot10^{-6} \tablenotemark{\bigstar}~; <2.1\cdot10^{-5} \tablenotemark{\bigstar}$ &
      & (3;4) &    &  \\
            & & & & & & & & $1.26\cdot10^{-5} \tablenotemark{\blacklozenge}~;<0.35\cdot10^{-6} \tablenotemark{\clubsuit}$ & & (5;6)& & \\
            & 33.1 & 3.00 & &(7)& & & & $8.6\cdot10^{-6} \tablenotemark{\blacklozenge}~$ & & (8)& & \\
            & 47.5& & &(9)& & & & $1.42\cdot10^{-5} \tablenotemark{\blacklozenge}$~;$1.1\cdot10^{-5} \tablenotemark{\blacklozenge}~$& & (10;9)& & \\
            & & & & & & & &  & 1880 &(11) & & \\
	    \hline
HD~64568    & 41 & 3.9 & 5.7 &(1)& 14 & 57 &(1)& 5.6$\cdot10^{-7}$ & 2800 &(1)& -7.5 &(1)\\
        &  &  &  & &  & & &  &  & &$<-5.97$ &(12) \\
	    \hline
HD~93250    & 40 & 4.0 & 5.5 &(1)& 12 & 53 &(1)& 5.5$\cdot10^{-7}$ & 2900 &(1)& -7.25 &(1) \\
        & 37.2 & & 5.75 &(13)& & 46.5 &(13)& & 3000 &(14)& -6.4 & (12)  \\
            & 52.5 & 3.95 & 6.4 &(15)& 19 & 120 &(15)&$ 1.3\cdot10{-6}$ & 3500 &(16)& & \\
            & 50.5 & 3.95/4.0\tablenotemark{b} & 6.28 &(17)& 18 & 118 &(17)&$4.9\cdot10^{-6}\tablenotemark{\blacklozenge}$ & 3250 &(17)& & \\
            & 51.0 & 3.9 & 6.3 &(18)& & 83.2/93.3/114.8\tablenotemark{e} &(18)&  & 3200;3230 &(18;11)& & \\
            & 46.0 & 3.95/3.96\tablenotemark{b} & 6.01 & (19) & 15.9 & 83.3 & (19) & $3.45\cdot10^{-6}$ $\tablenotemark{\blacklozenge}$& &(19) & & \\
            & 55.0 &  & &(9)& & & &$4.1\cdot10^{-5} \tablenotemark{\bigstar}~;7.9\cdot10^{-8}$ & &(20;21)& & \\
	    \hline
HD~93205    & 40 & 4.0 & 5.3 &(1)& 9 & 30 &(1)& 3.3$\cdot10^{-7}$ & 3200 &(1)& -7.0 &(1)\\
            & 55.0/36.5\tablenotemark{c} & & &(22) &12.5/10.3 \tablenotemark{c,f}&$\geq$39 / $\geq$15\tablenotemark{c}  &(22) & $>5.9\cdot10^{-7}$ &3600 & (23) & -6.37 & (23) \\
            & 44.7/35.5\tablenotemark{c} & &6.20/4.91\tablenotemark{c} &(24) & &100.00 / 24.82\tablenotemark{c}  &(24) &  & & & & \\
            & 36.3 & & 6.02 &(13) & & 63.4 &(13) &  & & & & \\
            & & &5.51/4.80\tablenotemark{c} &(25) & 8.0/6.3\tablenotemark{c} & 45/20\tablenotemark{c} &(25) &  & & &-6.45 &(12) \\
            &55.0 & & &(9) & &$\geq$31.5 / $\geq$13.3\tablenotemark{c} &(26) &  & & &$\sim$-7.0 &(26) \\
            & & & & & & $\geq$37/$\geq$15\tablenotemark{c} &(27) &  & & & & \\
            & & & & & & $\geq$32.6 / $\geq$14.2\tablenotemark{c} &(28) &  & & & & \\
            & & & & & & 63.3 / 24.5\tablenotemark{c}~; $\leq$60/$\leq$25\tablenotemark{c} &(29;30) &  & & & & \\
	    \hline
HDE~303308  & 40 & 4.0 & 5.3 &(1)& 9 & 30 &(1)& 3.4$\cdot10^{-7}$ & 2800 &(1)& -7.0 &(1)\\
        & 48.0 & 4.05/4.10\tablenotemark{b} & 5.84 &(17)& 12 & 66 &(17)& $2.1\cdot10^{-6}\tablenotemark{\blacklozenge}$ & 3100 &(17)& -6.54 & (12)\\
            & 45.5 & 3.9 & 5.8 &(31)&13&50/60/70\tablenotemark{d} &(31) & $2.5\cdot10^{-6}$ & 3400 &(16)& & \\
            & 35.5 & & 5.27 &(13) & & 28.8 &(13) & & 3035 & (11) & & \\
            & 48.0 & 3.9 & 5.85 &(18)& &40.7/41.7/40.7\tablenotemark{e} &(18)&  &3200;3000 &(18;14)& & \\
            & 41.0 & 3.90/3.91\tablenotemark{b} & 5.53 & (19) & 11.5 & 39.0 & (19) & $1.63\cdot10{-6}$$\tablenotemark{\blacklozenge}$& &(19) & & \\
            & 55.0& & &(9) & & & &  & & & & \\
	    \hline
HD~96715    & 39 & 4.0 & 5.2 &(1)& 9 & 30 &(1)& 2.2$\cdot10^{-7}$ & 2800 &(1)& -7.25 &(1)\\
        &  &  &  & &  &  & &  &2900;3000 &(14;11) &$<-5.85$ &(12) \\
	    \hline
HD~168076   & 39 & 4.0 & 5.2 &(1)& 9 & 30 &(1)& 2.2$\cdot10^{-7}$ & 2800 &(1)& -7.0 &(1)\\
        & 55.0 & 4.3 &  &(32) &  &  & &$2.51\cdot10^{-6}$\tablenotemark{\blacklozenge}& &(5) & & \\
            & 50.0 & & &(9)& & & & $1\cdot10^{-5}\tablenotemark{\blacklozenge}$ & &(9) & & \\
            & & & & & & & &  & 3305 &(11) & & \\
\enddata

\tiny{
\tablenotetext{a}{Mass-loss rate calculated from line fit to IUE lines 
unless a different method is indicated
with one of the following symbols:
$\bigstar $ analysis of radio data,
$\blacklozenge $ analysis of H$\alpha$,
$\clubsuit $ analysis of IR data.}
\tablenotetext{b}{
\logg derived from the fit to H$\gamma$ / Same but corrected for centrifugal force.}
\tablenotetext{c}{ Main
component / Secondary component.}
\tablenotetext{d}{ M/\Msun~
calculated from:
\logg and radius /
comparison 
with evolutionary tracks of high mass
loss rate /
same with tracks
of low mass-loss rate.}
\tablenotetext{e}{ M/\Msun~
calculated from: 
\vinf vs \Mstar relation / 
\logg and radius / comparison with evolutionary tracks.}
\tablenotetext{f}{ 
Assuming a distance of 2.6 Kpc to Tr16.}
}

\tablerefs{
\tiny{
(1) This work,
(2) \citet[]{SJ95}, (3) \citet[]{Sal98}, (4) \citet[]{ABC80}, (5) \citet[]{L88},
(6) \citet[]{PFG83}, (7) \citet[]{oke}, (8) \citet[]{Sal92}, (9) \citet[]{C77},
(10) \citet[]{MPRM03}, (11) \citet[]{Hal97},
(12) \citet[]{chs89}, (13) \citet[]{KG93}, (14) \citet[]{LSL95},
(15) \citet[]{K80}, (16) \citet[]{Gal81}, (17) \citet[]{puls96}, (18) \citet[]{Kal92},
(19) \citet[]{RPH03}, (20) \citet[]{Lal95}, (21) \citet[]{CG80}, (22) \citet[]{CW76},
(23) \citet[]{CG91}, (24) \citet[]{VG82}, (25) \citet[]{Aal00}, (26) \citet[]{Mal01},
(27) \citet[]{Cal99}, (28) \citet[]{SL93}, (29) \citet[]{DL84}, 
(30) \citet[]{Bal02}, (31) \citet[]{Sal83}, (32) \citet[]{M75}
}
}

\end{deluxetable}
}

\clearpage



\clearpage

\newpage

\clearpage

\newpage

\clearpage

\clearpage

\newpage

\clearpage

\newpage

\clearpage

\newpage

\clearpage

 \end{document}